\documentclass[a4paper,fleqn,usenatbib,useAMS]{mnras}

\usepackage{newtxtext,newtxmath}

\usepackage[T1]{fontenc}
\usepackage{ae,aecompl}

\usepackage{graphicx}	
\usepackage{amssymb}	
\usepackage{multicol}        
\usepackage{bm}		
\usepackage{enumitem}

\title[BRITE photometry of V973~Scorpii]{A \emph{BRITE}\thanks{Based on data collected by the \emph{BRITE-Constellation} satellite mission, designed, built, launched, operated and supported by the Austrian Research Promotion Agency (FFG), the University of Vienna, the Technical University of Graz, the Canadian Space Agency (CSA), the University of Toronto Institute for Aerospace Studies (UTIAS), the Foundation for Polish Science \& Technology (FNiTP MNiSW), and National Science Centre (NCN).} view on the massive O-type supergiant V973~Scorpii: Hints towards internal gravity waves or subsurface convection zones}

\author[Ramiaramanantsoa et al.]{Tahina Ramiaramanantsoa,$^{1,2}$\thanks{E-mail: tahina@astro.umontreal.ca} Rathish Ratnasingam,$^{3}$ Tomer Shenar,$^{4}$ \newauthor Anthony F. J. Moffat,$^{1,2}$ Tamara M. Rogers,$^{3,5}$ Adam Popowicz,$^{6}$ Rainer Kuschnig,$^{7}$ \newauthor Andrzej Pigulski,$^{8}$ Gerald Handler,$^{9}$ Gregg A. Wade,$^{10}$ Konstanze Zwintz,$^{11}$ \newauthor Werner W. Weiss$^{7}$  
\\\\
$^{1}$ D\'epartement de physique, Universit\'e de Montr\'eal, CP 6128, Succursale Centre-Ville, Montr\'eal, Qu\'ebec, H3C 3J7\\
$^{2}$ Centre de Recherche en Astrophysique du Qu\'ebec (CRAQ), Canada\\
$^{3}$ Department of Mathematics and Statistics, Newcastle University, Newcastle upon Tyne NE1 7RU, UK\\
$^{4}$ Institut f\"{u}r Physik und Astronomie, Universit\"{a}t Potsdam, Karl-Liebknecht-Str. 24/25, D-14476 Potsdam, Germany\\
$^{5}$ Planetary Science Institute, Tucson, AZ 85721, USA\\
$^{6}$ Instytut Automatyki, Politechnika \'Sl\c{a}ska, Akademicka 16, 44-100 Gliwice, Poland\\
$^{7}$ Institut f\"{u}r Astrophysik, Universit\"{a}t Wien, T\"{u}rkenschanzstrasse 17, 1180 Wien, Austria\\
$^{8}$ Instytut Astronomiczny, Uniwersytet Wroc{\l}awski, Kopernika 11, 51-622, Wroc{\l}aw, Poland\\
$^{9}$ Nicolaus Copernicus Astronomical Center, Bartycka 18, PL-00-716 Warsaw, Poland\\
$^{10}$ Department of Physics and Space Science, Royal Military College of Canada, Kingston, ON K7K 7B4, Canada\\
$^{11}$ Institut f\"{u}r Astro- und Teilchenphysik, Universit\"{a}t Innsbruck, Technikerstrasse 25, A-6020 Innsbruck, Austria\\
}

\date{Accepted 2018 July 12. Received 2018 June 30; in original form 2018 April 22}

\pubyear{2017}

\begin{document}
\label{firstpage}
\pagerange{\pageref{firstpage}--\pageref{lastpage}}
\maketitle

\begin{abstract}
Stochastically-triggered photospheric light variations reaching $\sim$$40$~mmag peak-to-valley amplitudes have been detected in the O8\,Iaf supergiant V973~Scorpii as the outcome of two months of high-precision time-resolved photometric observations with the \emph{BRIght Target Explorer (BRITE)} nanosatellites. The amplitude spectrum of the time series photometry exhibits a pronounced broad bump in the low-frequency regime ($\lesssim0.9$~d$^{-1}$) where several prominent frequencies are detected. A time-frequency analysis of the observations reveals typical mode lifetimes of the order of $5-10$~days. The overall features of the observed brightness amplitude spectrum of V973~Sco match well with those extrapolated from two-dimensional hydrodynamical simulations of convectively-driven internal gravity waves randomly excited from deep in the convective cores of massive stars. An alternative or additional possible source of excitation from a subsurface convection zone needs to be explored in future theoretical investigations.
\end{abstract}

\begin{keywords}
stars: massive --- stars: supergiants --- Physical Data and Processes: waves --- Physical Data and Processes: convection --- techniques: photometric
\end{keywords}



\section{Introduction}
\label{sec:V973Sco_Intro}

The stellar demographics of galaxies are such that massive O-type stars are heavily outnumbered by long-lived cool low-mass M dwarfs \citep[e.g.][]{1955ApJ...121..161S,2001MNRAS.322..231K,2010ARA&A..48..339B}. However, hot luminous stars more than compensate for their sparse spatial density by simply constituting the indispensable recycling engines of chemical elements involved in the formation of new stars and new substellar objects. That occurs both through their termination as supernovae and through the production of elements heavier than iron from the kilonovae occasionally induced by binary neutron star mergers \citep[e.g.][]{2017Sci...358.1570D,2017Natur.551...80K}. Thus, understanding the physical properties and the structures of massive stars at all their evolutionary stages prior to their deaths is of high interest. To that end, observational endeavours to probe their intrinsic variability (e.g. pulsations, rotational modulations, stochastic variations, episodic events such as flaring) -- which can yield constraints on their internal structures \citep[see e.g.][]{2010aste.book.....A} -- have been energized by the outcome of long-term, high-quality, high-cadence monitoring provided by space-based observatories (see \citeauthor{2018MNRAS.473.5532R}~\citeyear{2018MNRAS.473.5532R} for a recent update on that subject in the particular case of O-type stars).

The late O-type luminous supergiant V973~Sco (HD~151804; Table~\ref{tab:V973Sco_StellarParams}) is one of the most luminous stars in the Sco OB1 association \citep{1978ApJS...38..309H}.  No signs of multiplicity have been found so far for V973~Sco \citep[e.g.][]{1977ApJ...215..561C}, putting it at the same rank as other single O-type stars like $\xi$~Per and $\zeta$~Pup for constituting good laboratories for probing the intrinsic variability of O-type stars. However, the fainter apparent magnitude of V973~Sco compared to those of $\xi$~Per and $\zeta$~Pup makes it less privileged when it comes to O-star variability studies.  

The first spectroscopic variability study of V973~Sco was performed by \citet{1977ApJ...217..760S}, who inspected two ultraviolet (UV) spectra of the star taken $\sim$$1.5$~yr apart and found no significant changes in the P Cygni profiles of its C~{\sc iii}~$\lambda1174-1176$ and N~{\sc v}~$\lambda\lambda1239, 1242$ resonance lines. In that regard, the UV resonance lines of V973~Sco are saturated, such that their absorption troughs do not exhibit variable recurrent blueward-propagating discrete absorption components \citep[DACs -- originally called `narrow absorption components';][]{1989ApJS...69..527H,1999A&A...344..231K}. These DACs are found to be nearly ubiquitous amongst O stars and are best interpreted as manifestations of the presence of large-scale corotating interaction regions in the stellar wind \citep{1984ApJ...283..303M,1996ApJ...462..469C}. Nevertheless, a possible occurrence of DACs has been discovered in the P Cygni profile of the He~{\sc i}~$\lambda5876$ optical line of V973~Sco from two spectroscopic variability investigations on the star: the $15$ optical spectra collected by \citet{1992ApJ...390..650F} over $\sim$$6.2$~days revealed signs of migrating absorption enhancements in the absorption trough of the He~{\sc i}~$\lambda5876$ line profile, variability features that were later also found in the more intensive but slightly shorter spectroscopic campaign led by \citet{1996A&A...311..264P} ($64$ spectra spread over $\sim$$5$~d) who attributed them to DACs, although the length of the campaign did not allow for a derivation of a possible DAC recurrence time-scale. In addition to these variations, the He~{\sc ii}~$\lambda4686$ wind-sensitive emission line of V973~Sco was found to be variable, showing no signs of periodicity but only fluctuations from one night to the next \citep{1983ApJ...271..691G}, a behaviour that could suggest a link to stochastic wind clumping activity. 

V973~Sco has also been the subject of some photometric variability studies conducted from the ground that revealed non-negligible light variability \citep{1989A&AS...79..263V,1992MNRAS.254..404B}. Particularly, the $\sim$$15$-d long Str\"{o}mgren~$b$ photometric monitoring of V973~Sco led by \citet{1992MNRAS.254..404B} revealed light variations having an rms scatter of $\sim$$15$~mmag and happening on time-scales of the order of days. These ground-based observations unfortunately evidently suffered from a limited time base, daily aliases, and large gaps. Here, we report on the outcome of an effort to better determine the nature of the photometric variability of V973~Sco through two months of high-cadence, high-precision optical photometric monitoring from space with the \emph{BRIght Target Explorer (BRITE)} mission.

\begin{table}
\caption{Stellar parameters for V973~Sco. $T_{\star}$ and $R_{\star}$ are evaluated at a Rosseland optical depth of $20$, whereas the effective temperature $T_{\rm eff}$ corresponds to a radius $R_{2/3}$ where the Rosseland optical depth is $2/3$.}
{\normalsize
\begin{center}
\begin{tabular}{l c c c}
\hline
\hline
 Parameter & & Value & Reference  \\
\hline
Spectral type		&					&	O8\,Iaf 					& (1)	\\
$V$				&					&	$5.232\pm0.014$			& (2)	\\
$B-V$			&					&	$0.066\pm0.013$			& (2)	\\
$U-B$			&					&	$-0.838\pm0.039$			& (2)	\\
$\varv_{\rm e}\sin i$ 	&[$\mathrm{km~s}^{-1}$]	&	$104 \pm 14$				& (3)	\\
$\varv_{\rm rad}$	&[$\mathrm{km~s}^{-1}$]	&	$-46.7 \pm15.0$			& (4)	\\
$\varv_{\rm tan}$	&[$\mathrm{km~s}^{-1}$]	&	$18.9 \pm13.5$				& (4)	\\
Distance 			&[kpc]				&	$1.91 \pm 0.57$			& (4)	\\
$\log(L/L_{\sun})$ 	&					&	$5.90 \pm 0.2$				& (5)	\\
$\log g$ 			&[$\mathrm{cm~s}^{-2}$]	&	$3.00 \pm 0.2$				& (5)	\\
$T_{\rm eff}$ 		&[kK]				&	$28.1 \pm 0.5$				& (5)	\\
$T_{\star}$ 		&[kK]				&	$29.0 \pm 0.5$				& (5)	\\
$R_{2/3}$			&[$R_{\sun}$]			&	$37.6 \pm 1.3$				& (5)	\\
$R_{\star}$		&[$R_{\sun}$]			&	$35.4 \pm 1.2$				& (5)	\\
$\dot{M}$			&[M$_{\sun}$Myr$^{-1}$] 	&	 $6.3\pm0.1$				& (5)	\\
$M$				&[M$_{\sun}$] 			&	 $40\pm5$				& (5)	\\
$\varv_{\infty}$ 		&[$\mathrm{km~s}^{-1}$] 		&	 $1445\pm100$				& (6)	\\ \hline
\end{tabular}
\end{center}}
{\footnotesize 
$^{(1)}$Galactic O-Star Catalog \citep{2014ApJS..211...10S}, $^{(2)}$\citet{2004ApJS..151..103M}, $^{(3)}$\citet{1997MNRAS.284..265H}, $^{(4)}$\citet{1998A&A...331..949M}, $^{(5)}$\citet{2009A&A...503..985C}, $^{(6)}$\citet{1990ApJ...361..607P}}
\label{tab:V973Sco_StellarParams}
\end{table}

\section{BRITE-Constellation photometry of V973~Scorpii}
\label{sec:V973Sco_Obs_BRITE}

The launch of the first \emph{BRITE} satellites in February 2013 marked the beginning of a generation of nanosatellites fully dedicated to astronomy. The mission rightly takes the name of \emph{BRITE-Constellation} as it currently consists of five operational nanosatellites: \emph{BRITE-Austria (BAb)}, \emph{UniBRITE (UBr)}, \emph{BRITE-Lem (BLb)}, \emph{BRITE-Heweliusz (BHr)}, and \emph{BRITE-Toronto (BTr)}, the small letter appended to each abbreviation indicating the passband in which each satellite operates (``b'' for a blue filter covering $390-460$~nm, and ``r'' for a red filter covering $545-695$~nm). Located in low-Earth orbits with orbital periods of the order of $100$~min, each of the $20\times20\times20$~cm \emph{BRITE} nanosatellites is equipped with a $3$-cm telescope feeding an uncooled $4008\times2672$-pixel KAI-11002M CCD, with a large effective unvignetted field of view of $24\degr\times20\degr$ to fulfil a single purpose: tracking the long-term photometric variability of bright stars ($V\lesssim6$) in two passbands, typically over $2-6$~months. Full technical descriptions of the mission were provided by \citet{2014PASP..126..573W} and \citet{2016PASP..128l5001P}.

\begin{table}
\caption{Characteristics of the \emph{BRITE} observations of V973~Sco. Quantities listed in the last four rows were assessed at post-decorrelation stage. The last entry $\sigma_{\rm rms}$ is the root mean square mean standard deviation per orbit.}
{\normalsize
\begin{center}
\begin{tabular}{l c}
\hline
\hline
 Parameter  & Value  \\
\hline
Satellite									&	\emph{BRITE-Heweliusz} 	\\
Start -- End dates	 [HJD-2451545.0]			&	$5654.543-5718.347$	\\
Observing mode$^{\star}$						&	Chopping				\\
Exposure time [s]							&	$2.5$				\\[4pt] 
Total number of data points					&	$17374$				\\
Number of points per orbit$^{\star\star}$			&	$30~[6-40]$			\\
Contiguous time per orbit$^{\star\star}$ [min]		&	$15.9~[2.3-18.7]$		\\
$\sigma_{\rm rms}$ [mmag]					&	$1.68$				\\ \hline
\end{tabular} 
\end{center}}
{\footnotesize $^{\star}$ See \citet{2016PASP..128l5001P} and \citet{2017A&A...605A..26P} for the descriptions of the different modes of observations for \emph{BRITE}.\\$^{\star\star}$ Median values. Bracketed values indicate the extrema.}
\label{tab:V973Sco_Obs_BRITE_Log}
\end{table}

\begin{figure*}
\includegraphics[width=18cm]{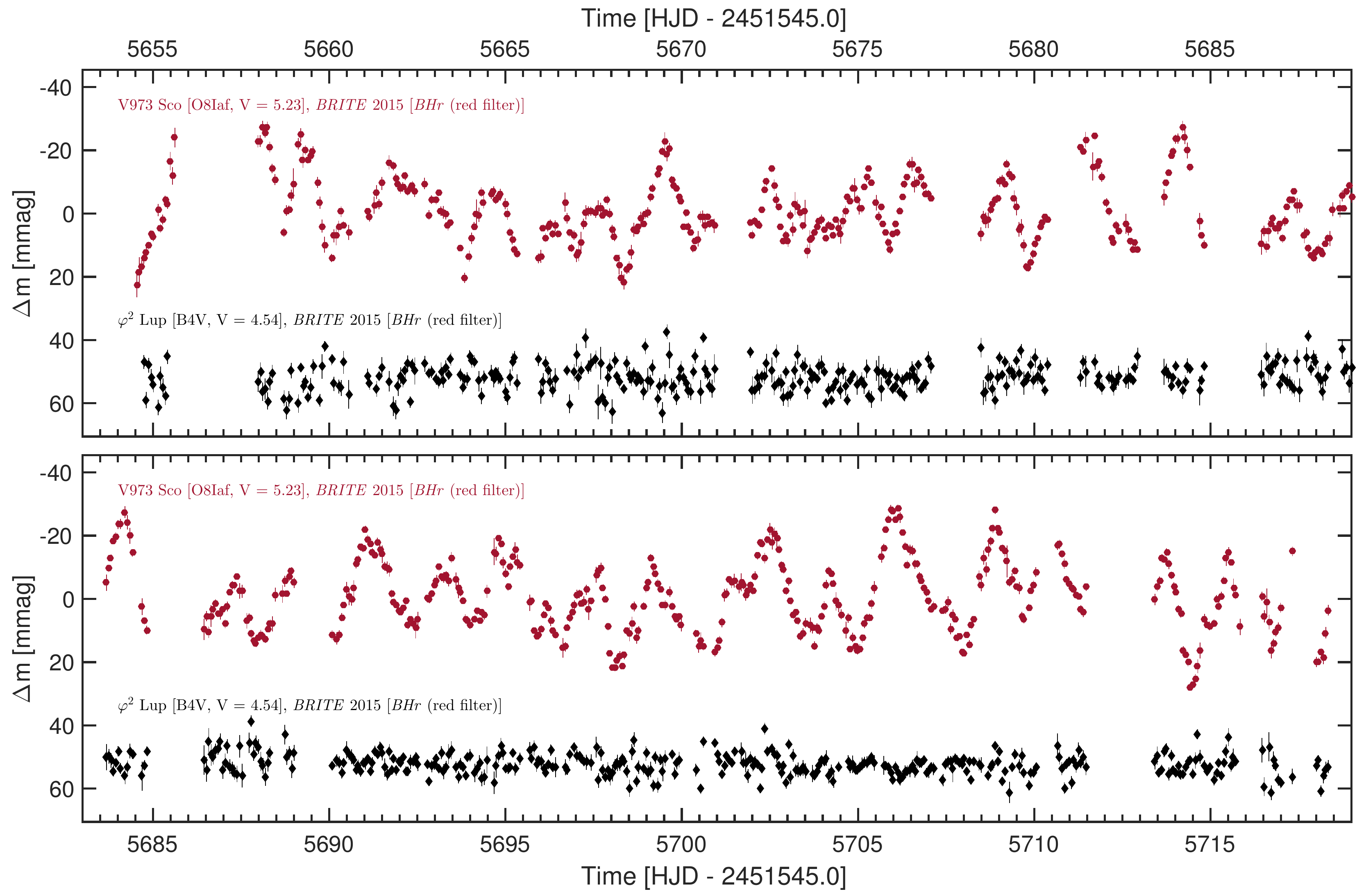}
 \caption{The two-month long \emph{BRITE} light curve of V973~Sco (red filled circles) as recorded in $2015$ in the red ($545-695$~nm) passband by \emph{BHr}, binned over each $\sim$$97.1$~min satellite orbit, along with the contemporaneous \emph{BHr} observations of $\varphi^2$~Lup (black diamonds, offset by $50$~mmag for better visibility), reduced and decorrelated with respect to instrumental effects in the same way as the light curve of V973~Sco but not showing any variability related to those observed in the latter (both in terms of amplitudes and time-scales). Small vertical bars indicate the $\pm1\sigma$ uncertainties. (Color versions of all figures in this paper are available in the online journal.)}
  \label{fig:V973Sco_BRITE_lcs_full}
\end{figure*}

\begin{figure}
\includegraphics[width=8.4cm]{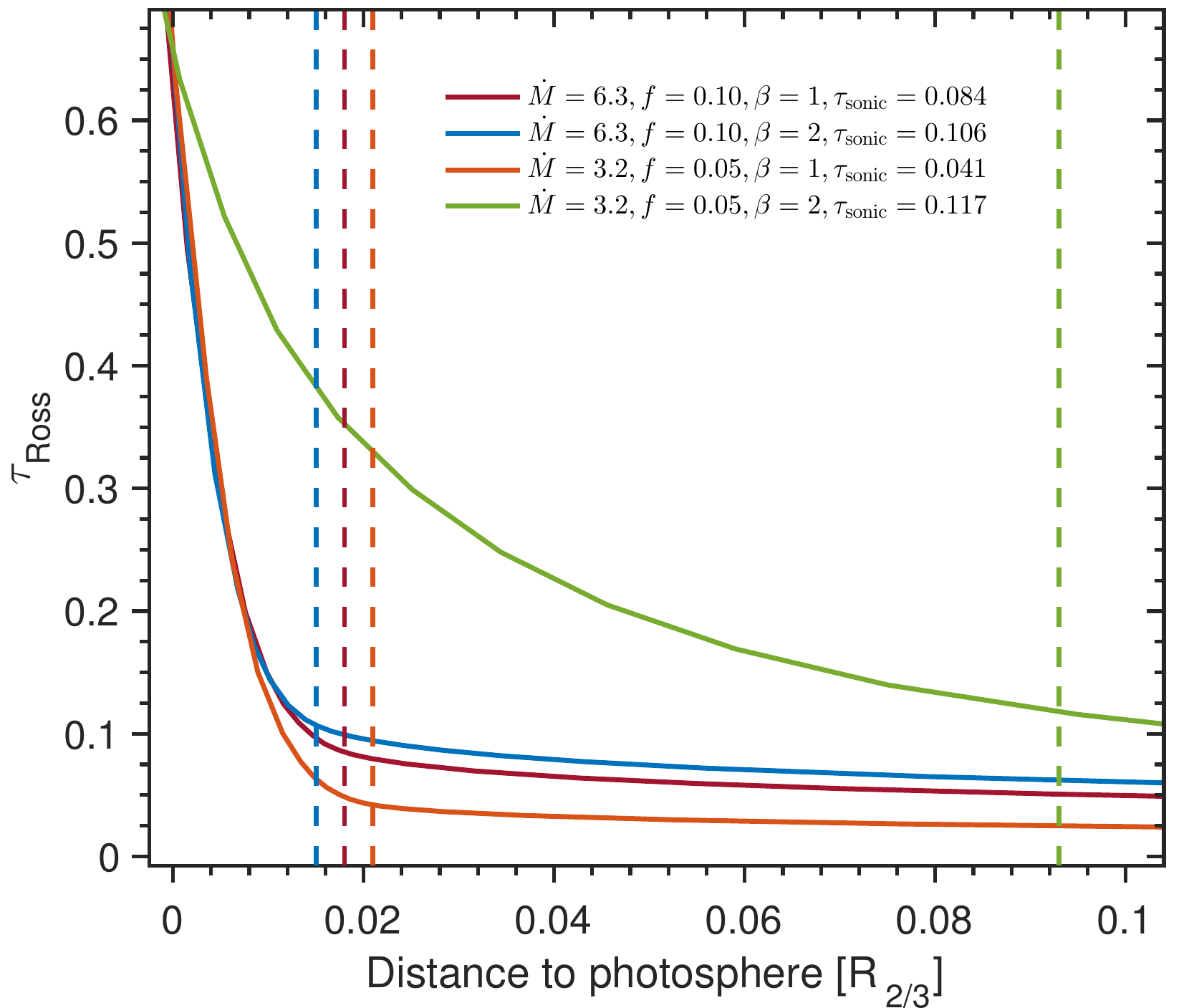}
 \caption{PoWR prediction of the evolution of the Rosseland mean optical depth with increasing distance from the photosphere ($\tau_{\rm Ross} = 2/3$), adopting the usual $\beta$-law velocity for the radial expansion of the wind, and considering various cases (see~Section~\ref{sec:V973Sco_Obs_BRITE}). For each case, the vertical dashed line indicates the limit beyond which the expansion becomes supersonic (i.e. the wind domain).}
  \label{fig:V973Sco_tauscale}
\end{figure}

During an observing run on the Scorpius field in 2015, \emph{BRITE} monitored a total of $26$ targets, only three of which are of spectral type O: $\mu$~Nor (O9.7\,Iab), V918~Sco (a detached massive O7.5\,If + ON9.7\,I binary) and V973~Sco. Four of the five operational \emph{BRITE} satellites were tasked to observe the field. \emph{BAb}, \emph{UBr} and \emph{BLb} were set to observe with the usual $1$~s exposure time, while \emph{BHr} was set to take longer $2.5$~s exposures to allow for the observation of fainter stars such as V918~Sco and V973~Sco with reasonable photometric precision. Although the entirety of the observing run on the field lasted $\sim$$5.5$~months (March 17, 2015 - August 29, 2015), V973~Sco was monitored by only one satellite, \emph{BHr}, for only about two months towards the end of the observing campaign, between June 26, 2015 and August 28, 2015 (HJD $2457199.543 - 2457263.347$), with individual $2.5$~s exposure snapshots taken at a median cadence of $27.3$~s during $\sim$$2-20\%$ of each $\sim$$97.1$~min orbit of \emph{BHr}. Table~\ref{tab:V973Sco_Obs_BRITE_Log} summarizes the characteristics of the \emph{BHr} observations of V973~Sco, the last four entries measured after cleaning the raw light curve from instrumental effects, a process that we describe in Appendix~\ref{sec:V973Sco_Decorrelation}. As illustrated in Fig.~\ref{fig:V973Sco_BRITE_lcs_full}, the \emph{BHr} light curve of V973~Sco exhibits clear variations reaching peak-to-valley amplitudes of $\sim$$40$~mmag. 

\section{How much does the stellar wind contribute to the observed light variations?}
\label{sec:V973Sco_PhotvsWind}

This has been investigated by \citet{2018MNRAS.473.5532R} in the case of the \emph{BRITE} dual-band observations of the hot early O-type supergiant $\zeta$~Puppis, for which it turned out that only a negligible fraction ($\lesssim0.35\%$) of the variability in the continuum could be due to a contribution from the wind, meaning that \emph{BRITE} essentially probed variations at the photosphere of $\zeta$~Pup. Although it is expected that the conclusion would not be very different in the case of our \emph{BRITE} observations of V973~Sco, we still redo the same analysis as in \citet{2018MNRAS.473.5532R} but with the stellar parameters of V973~Sco. To this end, we assessed the fraction of scattered photons in the stellar wind by constructing a PoWR \citep[Potsdam Wolf-Rayet;][]{2003A&A...410..993H,2004A&A...427..697H,2015A&A...579A..75T,2015ApJ...809..135S} model atmosphere corresponding to the stellar parameters of V973~Sco \citep[$L$, $g$, $T_{\rm eff}$, $\dot{M}$, and $v_{\infty}$ as listed Table~\ref{tab:V973Sco_StellarParams}, all from the investigation led by][]{2009A&A...503..985C}, then plotting the variation of the Rosseland mean optical depth with respect to the distance from the photosphere. We note that PoWR assumes the usual $\beta$-law velocity generally adopted in hot luminous stars for the radial component of the wind expansion. However, there is no explicit empirically-derived value of the exponent $\beta$ available for the case of V973~Sco. \citet{2009A&A...503..985C} found that $\beta=2$ yielded an excellent fit to the H$\alpha$ profile of ${\rm He}~3$--$759$ (O8\,Iaf), which shares very similar properties with V973~Sco. Therefore, we run the PoWR simulations for the cases $\beta=2$ and $\beta=1$, the latter being found to be more common for the winds of O-type stars. Additionally, we take wind clumping into account, considering the case of the standard assumption of a clumping filling factor $f=0.1$ \citep[as adopted by][]{2009A&A...503..985C}, and $f=0.05$ \citep[as adopted for $\zeta$~Pup;][]{2018MNRAS.473.5532R}. Finally, we noticed that the existing estimates of the mass loss rate for V973~Sco \citep[$12$~M$_{\sun}$Myr$^{-1}$ without clumping corrections, and $6.3$~M$_{\sun}$Myr$^{-1}$ when accounting for clumping;][]{2009A&A...503..985C} are curiously larger than typical mass loss rates for O-type stars (even $\zeta$~Pup has $\dot{M}=1.9$~M$_{\sun}$Myr$^{-1}$), such that we consider the cases $\dot{M}=6.3$~M$_{\sun}$Myr$^{-1}$ and $\dot{M}=3.2$~M$_{\sun}$Myr$^{-1}$. Under these considerations, the outcome of our simulations is illustrated in Fig.~\ref{fig:V973Sco_tauscale}, with the values of the resulting wind optical depth indicated for each case. Thus, the worst-case scenario points toward a wind that scatters $1-e^{-\tau_{\rm sonic}}\sim$$11\%$ of the photons coming from the photosphere. Since the optical wind-sensitive lines of V973~Sco are known to be variable at $\lesssim$10$\%$ of the continuum level \citep[e.g.][]{1996A&A...311..264P}, we can infer that a maximum of $\sim$$1.1\%$ of the variability in the continuum could arise from the wind. Hence, \emph{BRITE} basically monitors the photosphere of V973~Sco. 

\section{Frequency analysis}
\label{sec:V973Sco_Results_Fourier}

\begin{figure*}
\includegraphics[width=18cm]{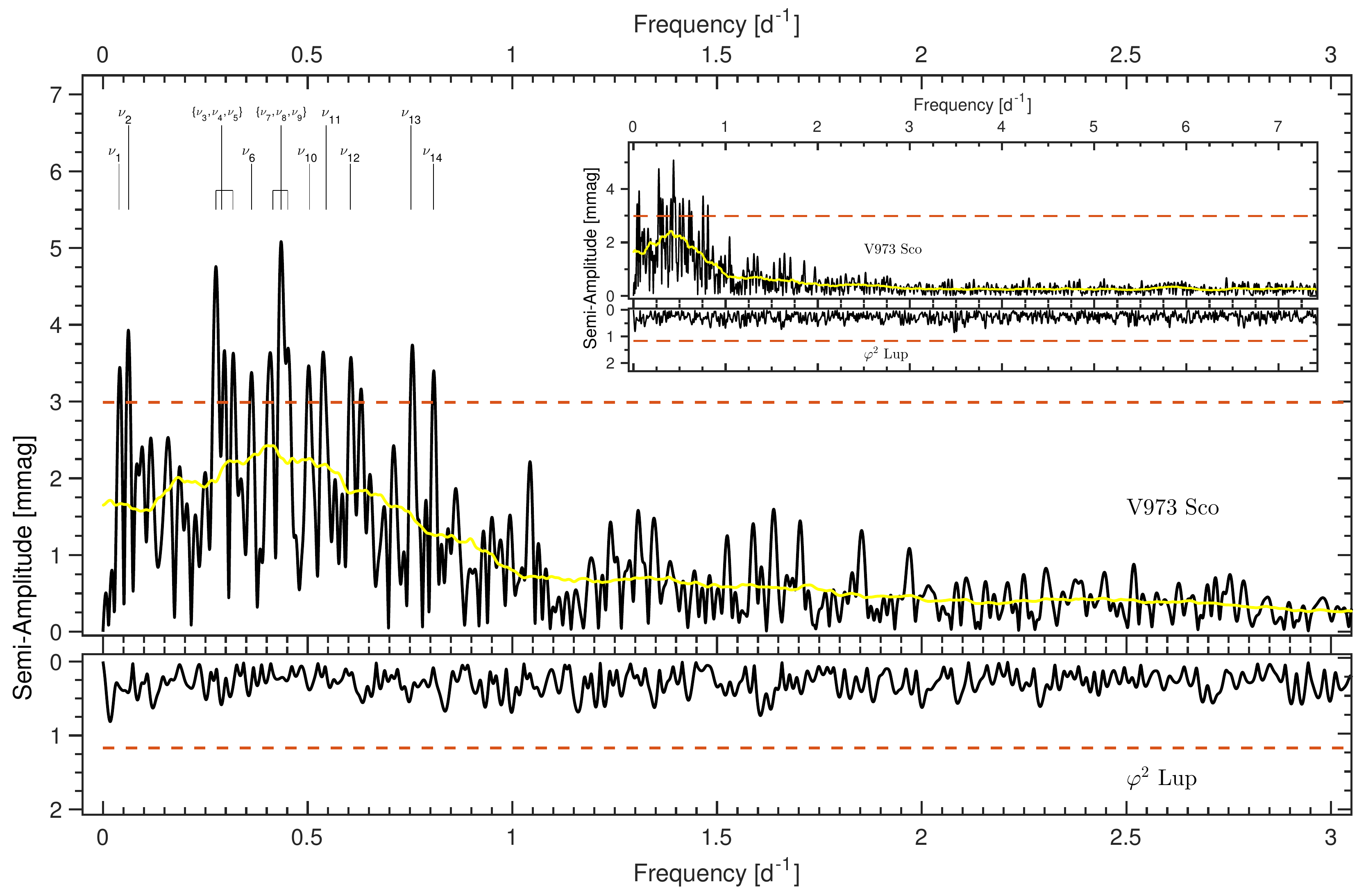}
 \caption{DFTs of the \emph{BHr} light curves of V973~Sco and $\varphi^2$~Lup, the dashed horizontal lines tracing four times the average noise levels evaluated over the frequency range $[0;4]$~d$^{-1}$. The inset zooms out on the overall behaviour up to the Nyquist frequency at $7.42$~d$^{-1}$. The DFT of the \emph{BHr} time series photometric observations of V973~Sco is illustrated along with its running average with a binwidth of $0.3$~d$^{-1}$, clearly revealing the bump in the low-frequency regime $[0;0.9]$~d$^{-1}$.}
  \label{fig:V973Sco_BRITE_DFT}
\end{figure*}

\begin{figure*}
\includegraphics[width=18cm]{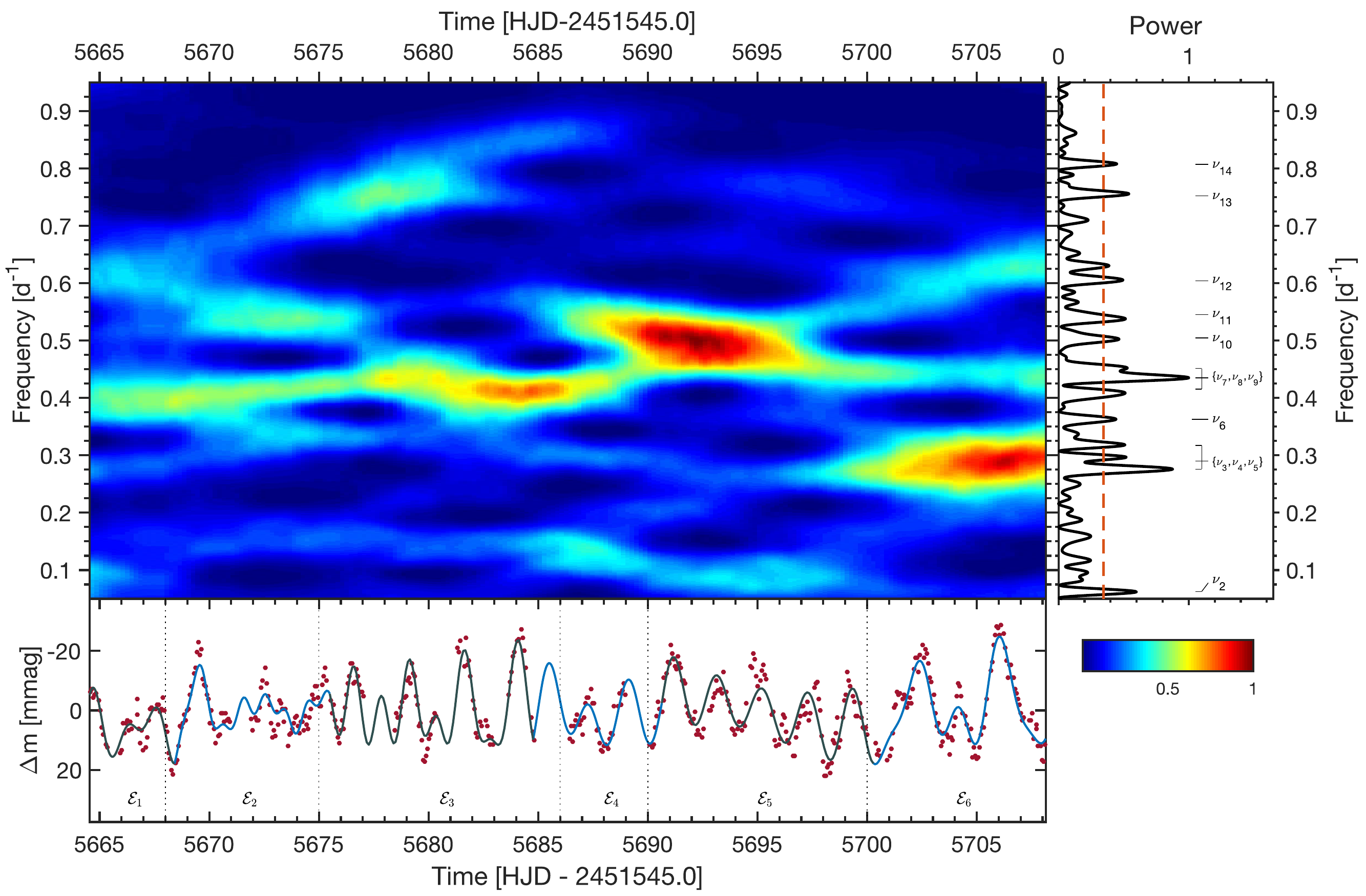}
 \caption{Time-frequency analysis of the \emph{BRITE} light curve of V973~Sco. \emph{Main panel:} spectrogram obtained from a $20$-d sliding tapered window Fourier transform of the light curve. \emph{Right panel:} Normalized power spectrum of the entire light curve, used as a guide in the determination of which frequencies are prominent during which epoch. The dashed line is the same as that in Fig.~\ref{fig:V973Sco_BRITE_DFT}, indicating four times the average noise level over $[0;4]$~d$^{-1}$. \emph{Bottom panel}: Light curve of V973~Sco (red points), with vertical dotted lines delimiting the epochs $\mathcal{E}_{m}$ obtained at the end of step (2) of the method described in Section~\ref{sec:V973Sco_Results_Fourier}. The continuous curves trace the final best fits obtained during each epoch.}
  \label{fig:V973Sco_BRITE_TFDiag}
\end{figure*}

To search for periodic signals in the light variations of V973~Sco, we evaluated the discrete Fourier transform (DFT) of the light curve with the {\sc period04} software package \citep{2005CoAst.146...53L} up to the Nyquist frequency ($7.42$~d$^{-1}$), resulting in the amplitude spectrum depicted in the upper panel of Fig.~\ref{fig:V973Sco_BRITE_DFT}. Immediately visible in this amplitude spectrum is a pronounced increase in amplitude towards low frequencies from $\sim$$1$~d$^{-1}$ to $\sim$$0.4$~d$^{-1}$, followed by a decrease below $\sim$$0.4$~d$^{-1}$, resulting in a bump in a confined low-frequency region where several prominent frequency peaks are present. In order to investigate the stability of these frequencies, we performed a time-frequency analysis by evaluating the windowed DFT of the light curve using a Hamming window of width $\mathcal{W} = 20$~d sliding through the time series at a step rate of $\mathcal{W} \times 1\%$, enabling us to cover up to $16$ cycles of the frequencies in the region of bumped amplitude at each step\footnote{Different window sizes were initially tested ($15$~d, $20$~d, $25$~d, and $30$~d). We chose $\mathcal{W} = 20$~d for the subsequent analyses as it offered the most reasonable compromise between the frequency resolution at each step, the number of signal cycles covered at each step, and the final time base covered by the spectrogram, which has to begin $\mathcal{W}/2$ after the first measurement in the time series and end $\mathcal{W}/2$ before the last observation.}. The main panel of Fig.~\ref{fig:V973Sco_BRITE_TFDiag} illustrates the resulting spectrogram, revealing random triggering of the prominent frequencies, which, when excited, are stable only for roughly $5-10$~days.  

\begin{table*}
\caption{The most prominent sine wave components of the intrinsic light variations of V973~Sco as observed by \emph{BRITE} in 2015. The first column indicates the labels for the parameters of each signal (frequencies $\nu_{n}$ in d$^{-1}$, amplitudes $A_n$ in mmag, and phases $\phi_n$ measured with respect to HJD-2451545.0 $ = 0.0$). The second column lists the preliminary frequency values extracted with {\sc Period04}, with the bracketed numbers indicating their S/N measured in the original DFT over $[0;4]$~d$^{-1}$. Columns $3-8$ list the subset epochs $\mathcal{E}_{m}$ of the observing run resulting from step (2) of the method described in Section~\ref{sec:V973Sco_Results_Fourier}, the bracketed numbers indicating their starting and ending dates (HJD-2451545.0). If a signal is prominent during epoch $\mathcal{E}_{m}$, its frequency, amplitude and phase resulting from the fitting procedure (step (3)) are listed in the corresponding column. The last line reports the standard deviation $\sigma_m$ (in mmag) of the residual light variations in each subset.}
\centering
{\scriptsize
\begin{center}
\begin{tabular}{c c c c c c c c c}
\hline
\hline
 \multicolumn{2}{c}{Signal}	&	 	&	\multicolumn{6}{c}{Epoch}		\\
\cline{1-2} \cline{4-9}\\[-5pt]

	&	&	&	$\mathcal{E}_1$	&	$\mathcal{E}_2$	&	$\mathcal{E}_3$	&	$\mathcal{E}_4$	&	$\mathcal{E}_5$	&	$\mathcal{E}_6$\\

	&	&	&	$[5663.04-5668.00]$	&	$[5668.00-5675.00]$	&	$[5675.00-5686.00]$	&	$[5686.00-5690.00]$	&	$[5690.00-5700.00]$	&	$[5700.00-5710.00]$\\

	&	 {\sc period04} &	&	&	&	&	&	\\

$\nu_1$	&	 0.03919 [4.6]&		&$-$	&$-$	&$-$	&$-$	&$-$	&$-$	\\
$A_1$	&	  $-$		  &		&$-$	&$-$	&$-$	&$-$	&$-$	&$-$	\\
$\phi_1$&		$-$	      &		&$-$	&$-$	&$-$	&$-$	&$-$	&$-$	\\\\

$\nu_2$	&	 0.06270 [5.3]&		&$-$	&$-$	&$-$	&$-$	&	$0.06225\pm0.00625$	&	$0.06278\pm0.00856$	\\
$A_2$	&$-$	 		      &		&$-$	&$-$	&$-$	&$-$	&	$5.35\pm0.63$	&	$3.61\pm0.59$	\\
$\phi_2$&$-$	 		      &		&$-$	&$-$	&$-$	&$-$	&	$^\star0.5039\pm0.0186$	&	$0.5039\pm0.0258$	\\\\

$\nu_3$	&	 0.27588 [6.4]&		&	$0.27597\pm0.01055$	&$-$	&$-$	&$-$	&$-$	&	$0.27594\pm0.00121$	\\
$A_3$	&$-$	 		      &		&	$8.49\pm0.67$	&$-$	&$-$	&$-$	&$-$	&	$25.55\pm0.59$	\\
$\phi_3$&$-$	 		      &		&	$0.2062\pm0.0125$	&$-$	&$-$	&$-$	&$-$	&	$0.1784\pm0.0036$	\\\\

$\nu_4$	&	 0.28999 [4.9]&		&$-$	&$-$	&$-$	&$-$	&$-$	&	$0.29043\pm0.00129$	\\
$A_4$	&$-$	 		      &		&$-$	&$-$	&$-$	&$-$	&$-$	&	$23.88\pm0.59$	\\
$\phi_4$&$-$	 		      &		&$-$	&$-$	&$-$	&$-$	&$-$	&	$0.0183\pm0.0039$	\\\\

$\nu_5$	&	 0.31742 [4.9]&		&$-$	&$-$	&$-$	&$-$	&$-$	&	$0.31746\pm0.00268$	\\
$A_5$	&$-$	 		      &		&$-$	&$-$	&$-$	&$-$	&$-$	&	$11.54\pm0.59$	\\
$\phi_5$&$-$	 		      &		&$-$	&$-$	&$-$	&$-$	&$-$	&	$0.3339\pm0.0081$	\\\\

$\nu_6$	&	 0.36287 [4.5]&		&	$0.36294\pm0.01180$	&$-$	&$-$	&$-$	&$-$	&$-$	\\
$A_6$	&	$-$		      &		&	$7.59\pm0.67$	&$-$	&$-$	&$-$	&$-$	&$-$	\\
$\phi_6$&	$-$		      &		&	$0.2005\pm0.0140$	&$-$	&$-$	&$-$	&$-$	&$-$	\\\\

$\nu_7$	&	 0.41538 [4.9]&		&	$0.41545\pm0.00493$	&	$0.39992\pm0.00348$&	$0.40918\pm0.00193$	&	$0.41534\pm0.00538$	&$-$	&$-$	\\
$A_7$	&	$-$			  &		&	$18.17\pm0.67$	&	$13.00\pm0.60$	&	$18.54\pm0.64$	&	$14.56\pm0.60$	&$-$	&$-$	\\
$\phi_7$&	$-$		      &		&	$0.2618\pm0.0058$	&	$0.2310\pm0.0074$	&	$0.0865\pm0.0055$	&	$0.0117\pm0.0066$	&$-$	&$-$	\\\\

$\nu_8$	&	 0.43498 [6.8] &	&$-$	&	$0.43888\pm0.00404$	&	$0.43385\pm0.00252$	&	$0.43536\pm0.01488$	&	$0.43479\pm0.00371$	&	$0.43508\pm0.00465$	\\
$A_8$	&	$-$ 		      &		&$-$	&	$11.20\pm0.60$	&	$14.19\pm0.64$	&	$5.26\pm0.60$	&	$9.02\pm0.63$	&	$6.64\pm0.59$	\\
$\phi_8$	&	$-$ 		      &		&$-$	&	$0.6858\pm0.0086$	&	$0.5448\pm0.0071$	&	$0.1056\pm0.0183$	&	$0.7952\pm0.0111$	&	$0.0000\pm0.0140$	\\\\

$\nu_9$	&	 0.45144 [4.9] &	&$-$	&$-$	&$-$	&	$0.45104\pm0.00417$	&	$0.45167\pm0.00319$	&$-$	\\
$A_9$	&	$-$		      &		&$-$	&$-$	&$-$	&	$18.79\pm0.60$	&	$10.46\pm0.63$	&$-$	\\
$\phi_9$	&$-$	 		      &		&$-$	&$-$	&$-$	&	$0.4847\pm0.0051$	&	$0.1533\pm0.0096$	&	$-$\\\\

$\nu_{10}$&	 0.50473 [4.6] &	&$-$	&$-$	&$-$	&$-$	&	$0.50434\pm0.00221$	&$-$	\\
$A_{10}$	 &	$-$ 		      &		&$-$	&$-$	&$-$	&$-$	&	$15.15\pm0.63$	&$-$	\\
$\phi_{10}$&$-$	 		      &		&$-$	&$-$	&$-$	&$-$	&	$0.4657\pm0.0067$	&$-$	\\\\

$\nu_{11}$&	 0.54549 [4.9] &	&$-$	&	$0.54553\pm0.01436$	&$-$	&	$0.54556\pm0.01081$	&	$0.54620\pm0.00554$	&	$-$\\
$A_{11}$	&$-$	 		      &		&$-$	&	$3.15\pm0.60$	&$-$	&	$7.24\pm0.60$	&	$6.03\pm0.63$	&	$-$\\
$\phi_{11}$&$-$	 		      &	  	&$-$	&	$0.0045\pm0.0305$	&$-$	&	$0.0079\pm0.0133$	&	$0.5354\pm0.0166$	&	$-$\\\\

$\nu_{12}$&	 0.60427 [4.8] &	&	$0.60419\pm0.00971$	&$-$	&$-$	&$-$	&$-$	&	$0.60387\pm0.00517$	\\
$A_{12}$	&	$-$ 		      &		&	$9.23\pm0.67$	&$-$	&$-$	&$-$	&$-$	&	$5.97\pm0.59$	\\
$\phi_{12}$&	$-$		      &	  	&	$0.2959\pm0.0115$	&$-$	&$-$	&$-$	&$-$	&	$0.1539\pm0.0156$\\\\

$\nu_{13}$&	 0.75239 [5.0] &	&$-$	&$-$	&	$0.75237\pm0.01194$	&$-$	&$-$	&$-$	\\
$A_{13}$	&	 $-$		      &		&$-$	&$-$	&	$3.00\pm0.64$	&$-$	&$-$	&$-$	\\
$\phi_{13}$&	 $-$		      &	  	&$-$	&$-$	&	$0.8084\pm0.0337$	&$-$	&$-$	&$-$	\\\\

$\nu_{14}$&	 0.80726 [4.6] &	&$-$	&$-$	&	$0.80235\pm0.00439$	&$-$	&$-$	&$-$	\\
$A_{14}$	&$-$	 		      &		&$-$	&$-$	&	$8.15\pm0.64$	&$-$	&$-$	&$-$	\\
$\phi_{14}$&$-$	 		      &	  	&$-$	&$-$	&	$0.1659\pm0.0124$	&$-$	&$-$	&$-$	\\\\

\cline{1-2} \cline{4-9}\\[-5pt]

\multicolumn{2}{c}{$\sigma_m$}	&	 	&	$3.47$	&	$4.21$	&	$4.33$	&	$3.12$	&	$5.24$	&	$4.83$	\\

\hline
\end{tabular}
{\scriptsize 
$^\star$Phase kept constant as the signal clearly started at the beginning of $\mathcal{E}_5$ and continues in $\mathcal{E}_6$ but with a lower amplitude.}
\end{center}}
\label{tab:V973Sco_FreqAnalysis}
\end{table*}

The next step is to quantify the characteristics of these prominent signals (their frequencies, amplitudes and phases). The standard way of achieving this is through the usual iterative prewhitening procedure using the entire time series. However, that procedure is efficient for the extraction of the properties of sinusoidal signals that are present  and relatively stable from the beginning until the end of the observations, but remains inappropriate for the characterization of stochastically excited signals with finite lifetimes. This simply comes from the fact that prewhitening a randomly triggered sinusoidal signal with finite lifetime using the entire time series would introduce a spurious signal with the same frequency at epochs during which the actual transient signal was not present. For instance, prewhitening the dominant frequency at $\sim$$0.5$~d$^{-1}$ present in the time interval $5690.0-5696.0$ in the light curve of V973~Sco would introduce a spurious sinusoidal signal at the same frequency in the time interval $5696.0-5710.0$. Additionally, such an approach might even partially or totally dilute some other signals at frequencies very close to the one being prewhitened. Alternatively, ``local prewhitening'' could be done only within subsets of the light curve where the signal is excited and stable enough. However, in that case, iterative prewhitening cannot be necessarily performed efficiently on a given subset, i.e. after prewhitening a signal that is prominent within a subset, one has to choose another subset to look for and characterize other prominent signals because those other signals may only be partly excited within the current subset. Moreover, since the first step of the prewhitening consists of computing the DFT to look for the first prominent signal, doing this on subsets of the light curve means dealing each time with amplitude spectra with poor frequency resolution. Under these considerations, we adopt the following three-step method to extract the characteristics of the transient sinusoidal signals present in the light curve of V937~Sco: 

\begin{enumerate}[labelindent=4.0pt,leftmargin=*,label={(\arabic*)}]
\item extraction of the values of the dominant frequencies present in the observations through the standard iterative prewhitening procedure using the entire light curve. We do this here \emph{only} to get first estimates of the values of the dominant frequencies, which we use as inputs for the subsequent steps;
\item determination of the epochs at which each of these signals are excited, followed by a splitting of the time series into subsets according to these epochs. This is performed visually using the information provided by the spectrogram, and the definition of the epochs are primarily driven by times where new signals appear;
\item multifrequency sine wave fit to the observed variations. For a given subset in which $N$ prominent sinusoidal signals are detected, we quantify the characteristics of each of these signals through a Levenberg-Marquardt non-linear least-squares fit of the model function $\sum_{n=1}^{N} A_n\sin\left[2{\upi}\left(\nu_n [t-t_0] + \phi_n\right)\right]$ to the observed variations in the subset. This sum describes the total signal arising from the contribution of the $N$ prominent sinusoidal signals triggered during the subset, each of frequency $\nu_n$, amplitude $A_n$, and phase $\phi_n$ with respect to the arbitrary fixed time reference $t_0$.      
\end{enumerate}

\begin{figure}
\includegraphics[width=8.4cm]{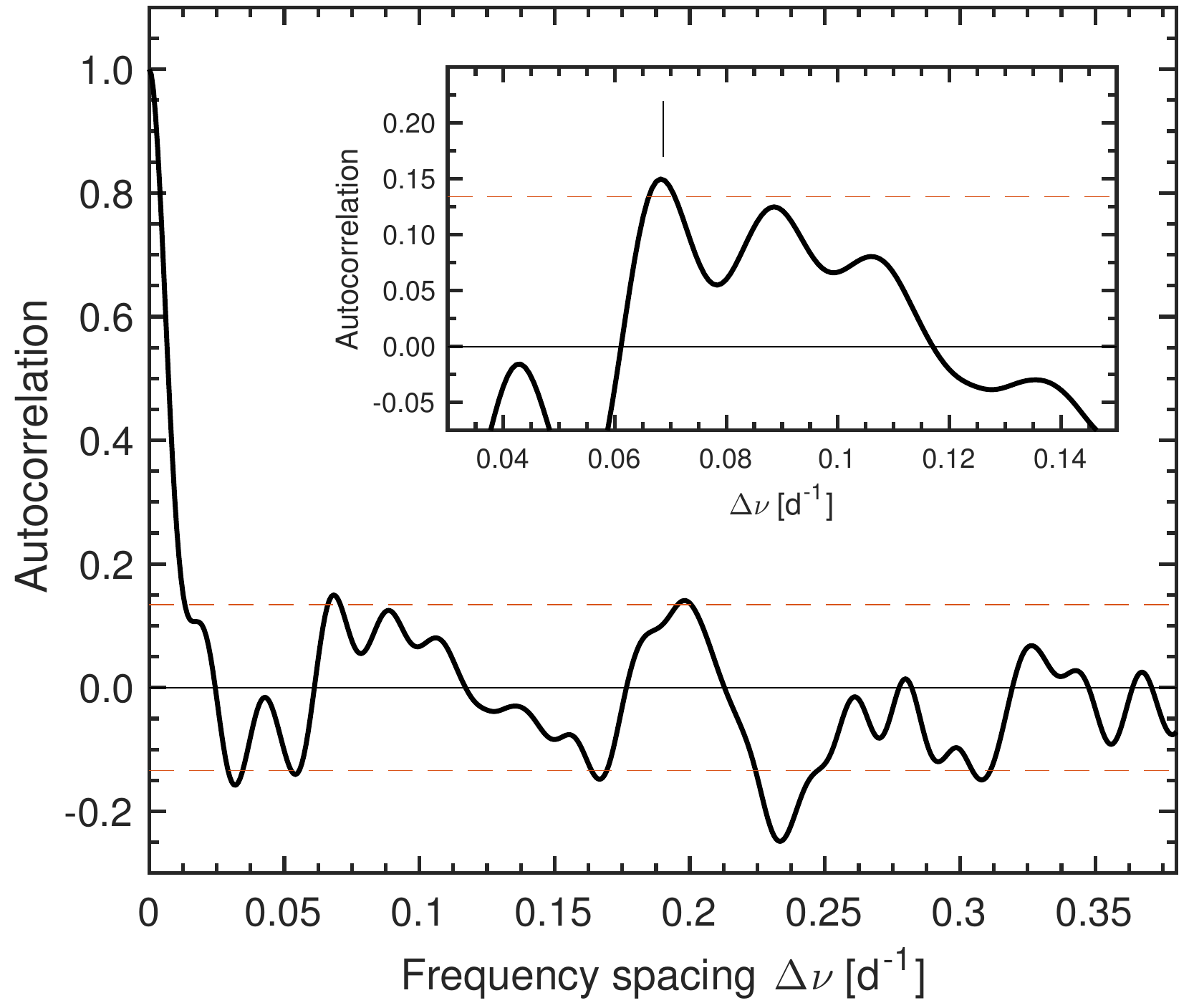}
 \caption{Autocorrelation function (ACF) of the $[0.2;0.9]$~d$^{-1}$ portion of the amplitude spectrum of the \emph{BRITE} observations of V973~Sco (continuous curve), along with the $4\sigma$ confidence bounds (dashed horizontal lines), the inset zooming on the region of lags $[0.03;0.15]$~d$^{-1}$ where a significant peak is detected at a lag of $\Delta\nu=0.069\pm0.006$~d$^{-1}$ corresponding to the frequency spacing apparently exhibited by the elements of set $\mathcal{S}_4=\left\{\nu_4,\nu_6,\nu_8,\nu_{10}\right\}$. The ACF also appears to present another peak at $\sim$$0.2$~d$^{-1}$ rising slightly above the $4\sigma$ threshold but having a relatively large width to be unambiguously considered a true spacing.}
  \label{fig:V973Sco_BRITE_DFT_ACF}
\end{figure}

The preliminary values of the prominent frequencies that we extracted with {\sc Period04} during step (1) are listed in the second column of Table~\ref{tab:V973Sco_FreqAnalysis} and indicated in Figs.~\ref{fig:V973Sco_BRITE_DFT}~and~\ref{fig:V973Sco_BRITE_TFDiag}. We note the detection of three sets of two--three unresolved frequencies, $\mathcal{S}_1=\left\{\nu_1,\nu_2\right\}$, $\mathcal{S}_2=\left\{\nu_3,\nu_4,\nu_5\right\}$ and $\mathcal{S}_3=\left\{\nu_7,\nu_8,\nu_9\right\}$, as $\left| \nu_1-\nu_2 \right| = 0.0235$~d$^{-1}$, $\left| \nu_3-\nu_4 \right| = 0.0141$~d$^{-1}$, $\left| \nu_5-\nu_4 \right| = 0.0274$~d$^{-1}$, $\left| \nu_7-\nu_8 \right| = 0.0196$~d$^{-1}$, and $\left| \nu_9-\nu_8 \right| = 0.0164$~d$^{-1}$, all below or barely above the \citet{1978Ap&SS..56..285L} threshold which in our case is $1.5/T\simeq0.0235$~d$^{-1}$ ($T$ being the total time base of the observations). Also, under this criterion (and even just considering the Rayleigh criterion), one frequency, $\nu_{11}=0.5047$~d$^{-1}$ [$P_{11}=1.9814$~d], could be the occurence of the first harmonic of $\nu_3=0.2759$~d$^{-1}$ [$P_3=3.6245$~d]. However, the possibility of this being purely coincidental cannot be excluded because there are a dozen prominent frequency peaks packed in the region $\sim$$[0.2;0.9]$~d$^{-1}$ and the probability of occurence of combinations increases with the number of detected frequencies. Finally, also as a result of the $1.5/T$ criterion, a set of four frequency peaks, $\mathcal{S}_4=\left\{\nu_4,\nu_6,\nu_8,\nu_{10}\right\}$, appears to share an equal spacing of $\sim$$0.07$~d$^{-1}$. This is further confirmed by the autocorrelation of the part of the amplitude spectrum in the range $\sim$$[0.2;0.9]$~d$^{-1}$ (Fig.~\ref{fig:V973Sco_BRITE_DFT_ACF}), revealing a frequency spacing of $\Delta\nu=0.069\pm0.006$~d$^{-1}$. However, again due to the high concentration of frequency peaks in the restricted $\sim$$[0.2;0.9]$~d$^{-1}$ region, pure coincidence cannot be ruled out for the origin of this spacing.

Then, in step (2) we ended up with the six epochs highlighted in the bottom panel of Fig.~\ref{fig:V973Sco_BRITE_TFDiag}, with typical time bases in the range $5-11$~days. Table~\ref{tab:V973Sco_FreqAnalysis} summarizes which signal is excited during which epoch. We note that the determination of the epochs of excitation of a prominent signal was not straightforward for some signals that do not have well-resolved frequencies, especially the two signals with frequencies $\nu_7$ and $\nu_8$. Wherever there is an ambiguity about whether such signals are excited or not, we decided to include them, and the result is that they appear to last longer than the other signals (however, their amplitudes clearly change from one epoch to another). Evidently, observations over a much longer time base are needed in the future to better assess the lifetimes of these signals. Speaking of time base, it should be kept in mind that, since the windowed Fourier transforms have to be evaluated starting $\mathcal{W}/2$ after the first observation in the time series and ending $\mathcal{W}/2$ before the last data point in the time series, the spectrogram is limited to that time base, and therefore this analysis is only based on epochs defined within that time span. Consequently, this approach is not adequate for the characterization of the very low frequency $\nu_1=0.0392$~d$^{-1}$ [$P_1=25.5102$~d]; but, in any case, since our observations only cover $\sim$$2.5$ cycles of this signal, further observations over a much longer time base are needed to validate whether this signal is real or an artifact of instrumental origin. Additionally, as already mentioned previously, the current data barely allow us to unambiguously distinguish $\nu_1$ and $\nu_2=0.0627$~d$^{-1}$ as two different frequencies. 

Finally, at step (3) we used the preliminary values of the frequencies determined in step (1) as first guesses for the frequencies involved in the fitting procedure. Particularly, in order to account for the \citet{1978Ap&SS..56..285L} criterion, we let the frequencies vary within $\pm1.5/T$ of their preliminary guesses, and leave the amplitudes and phases as completely free parameters to be determined. Furthermore, we note another key point that has to be taken care of: continuity at subset boundaries. More precisely, ideally the transitions at the boundaries need to result in an overall function of class at least $C^1$ (i.e. differentiable and having a continuous derivative). In order to account for this subtle issue, we adopt the following strategy:  

\begin{enumerate}[labelindent=4.0pt,leftmargin=*,label={(3.\arabic*)}]
\item perform the fitting procedure on the different subsets independently;
\item note the subset $\mathcal{E}_{\rm best}$ which yielded the best fit (highest adjusted R$^2$ statistic);
\item consider an adjacent subset $\mathcal{E}_{{\rm adj},1}$ and slightly extend its time base to overlap with that of $\mathcal{E}_{\rm best}$, thus obtaining an extended subset $\mathcal{E}^{\rm ext}_{{\rm adj},1}$;
\item approximate the variations in the overlapping zone by the fitted variations from $\mathcal{E}_{\rm best}$, and redo the fit in the extended subset $\mathcal{E}^{\rm ext}_{{\rm adj},1}$;
\item the best fit in subset $\mathcal{E}^{\rm ext}_{{\rm adj},1}$ is therefore taken to be the one that has a high-enough adjusted R$^2$ value and is such that the transition to the fitted variations in subset $\mathcal{E}_{\rm best}$ results in a nearly class $C^1$ function;
\item propagate the procedure in adjacent subsets until all epochs are treated.       
\end{enumerate}

At the end of step (3.1), it turned out that $\mathcal{E}_{\rm best}=\mathcal{E}_6$ (R$^2\sim$$82\%$), such that we essentially went backwards in time to treat the subsequent adjacent epochs. The overlaps typically span around $0.5-0.75$~d depending on the gaps in the observations, with an exception of a $2$-day overlap between subsets $\mathcal{E}_4$ and $\mathcal{E}_3$ due to the relatively large $\sim$$1.6$-day gap in the observations around this transition, and for which we failed to determine a reasonable behaviour of class $C^1$, as we only found a continuous but not formally differentiable behaviour. The final fits to the observed variations in the subsets have R$^2$ values in the range $\sim$$71-82\%$ and are depicted in the bottom panel of Fig.~\ref{fig:V973Sco_BRITE_TFDiag}. The best-fit parameters for each sine wave component in each subset are listed in columns $3-8$ of Table~\ref{tab:V973Sco_FreqAnalysis}. We also quote the analytical $1\sigma$ uncertainties for these signal parameters \citep{1999DSSN...13...28M}:
\begin{equation}
\begin{cases}
~\sigma(\nu_{n,m}) = \sqrt{\frac{6}{N_m}} \frac{1}{\upi T_m} \frac{\sigma_{m}}{A_{n,m}}, \\[5pt]
~\sigma(A_{n,m}) = \sqrt{\frac{2}{N_m}}\sigma_{m},\\[5pt]
~\sigma(\phi_{n,m}) = \frac{1}{2\upi} \sqrt{\frac{2}{N_m}} \frac{\sigma_{m}}{A_{n,m}},
\end{cases}
\label{eq:V973Sco_FormalErrors}
\end{equation}   
where $N_m$ is the number of observations during the time span $T_m$ of epoch $\mathcal{E}_m$. In the original formulation of these analytical uncertainties, $\sigma_{m}$ represents the root mean square error in the observations. It has to be kept in mind that the resulting formal uncertainties on the signal parameters are known to underestimate the true uncertainties \citep[see e.g.][]{2000MNRAS.318..511H}. Here we even adopt a more conservative approach by considering $\sigma_{m}$ to be the standard deviation of the residual time series after removal of the fitted multiperiodic signal \citep[][p. 367--368]{2005CoAst.146...53L,2010aste.book.....A}, which is typically a factor $\sim$$2-3$ higher than the root mean square errors in the observations during each epoch $\mathcal{E}_m$. An even more pessimistic approach could consist in directly adopting those $\sigma_m$ values (also indicated in Table~\ref{tab:V973Sco_FreqAnalysis}) as gross estimates of the uncertainties in the signal amplitudes.

\section{Origin of the light variations}
\label{sec:V973Sco_Results_Interpretation}

First of all, the fact that $\nu_{11}\simeq2\nu_3$ within the limits set by the \citet{1978Ap&SS..56..285L} criterion (and even just considering the Rayleigh criterion) is noteworthy. Interpreting $\nu_3=0.2759$~d$^{-1}$ [$P_3=3.6245$~d] as the rotational frequency $\nu_{\rm rot}$ is not evident at all and is highly unlikely because it corresponds to an equatorial velocity very close to the critical rotational velocity. Indeed, adopting $\varv_{\rm crit}=\sqrt{\mathcal{G}M/R_{\rm e}}$ \citep[$M$ being the stellar mass, $R_{\rm e}$ the radius at the equator, and $\mathcal{G}$ the gravitational constant; e.g.][]{2004MNRAS.350..189T}, which can be simply re-written as $\varv_{\rm crit}=\sqrt{gR_{\rm e}}$ (since $M=gR_{\rm e}^2/\mathcal{G}$), and taking into account the uncertainties on the values available to date for the surface gravity and radius of V973~Sco (Table~\ref{tab:V973Sco_StellarParams}), the critical rotational velocity would be at $511_{-112}^{+143}$~km~s$^{-1}$, corresponding to a critical rotational frequency of $0.2689_{-0.0589}^{+0.0756}$~d$^{-1}$. While this sets an upper limit on $\nu_{\rm rot}$, its lower bound can be obtained from the estimate of the projected rotational velocity available to date. In fact, the minimum value that $\nu_{\rm rot}$ can reach is $\varv_{\rm e}\sin i/(2\upi R_{\rm e}) = 0.0547_{-0.0089}^{+0.0096}$~d$^{-1}$ (imposed by $\sin i \leq 1$). Hence, under the \citet{1978Ap&SS..56..285L} criterion, $\nu_2 = 0.06270$~d$^{-1}$ [$P_2=15.9490$~d] would be a better candidate for the rotational frequency (in which case V973~Sco would be seen almost equator on). However, as already repeatedly noted in Section~\ref{sec:V973Sco_Results_Fourier}, the two frequencies in set $\mathcal{S}_1=\left\{\nu_1,\nu_2\right\}$ are barely distinguishable. Again, observations over a longer time base are needed in order to check this possibility.  

Secondly, as already mentioned in Section~\ref{sec:V973Sco_Results_Fourier}, we noticed the presence of a regular spacing of $\Delta\nu=0.069\pm0.006$~d$^{-1}$ between the elements of set $\mathcal{S}_4$. Three possibilities are to be explored for the origin of this frequency spacing:

\begin{enumerate}[labelindent=4.0pt,leftmargin=*,label={(\arabic*)}]
\item large frequency separation exhibited by adjacent pressure modes of the same low degree but different high overtone in the asymptotic approximation \citep[][]{1980ApJS...43..469T,1990ApJ...358..313T};
\item small frequency separation of $p$ modes predicted by the asymptotic theory;
\item rotational splitting of pressure modes of the same degree and same overtone but different azimuthal order \citep{1951ApJ...114..373L}.
\end{enumerate}

Regarding (1), for a star as massive and large as V973~Sco, scaling laws \citep{1995A&A...293...87K} yield a large frequency separation of about a factor five greater than $\Delta\nu$, meaning that this spacing is unlikely to be a large frequency separation. Then remain possibilities (2) and (3). First of all, if $\Delta\nu$ is related to rotational splitting of $p$ modes, only three adjacent members of set $\mathcal{S}_4$ would be involved, i.e. either $\mathcal{S}^{\prime}_4=\left\{\nu_4,\nu_6,\nu_8\right\}$, or $\mathcal{S}^{\prime\prime}_4=\left\{\nu_6,\nu_8,\nu_{10}\right\}$.  \emph{However}, before going further into these interpretations, the first question that needs to be answered is: can the relatively low frequencies involved in set $\mathcal{S}_4$ arise from pressure modes for an evolved hot massive star like V973~Sco? In that regard, it should first be pointed out that the stellar parameters of V973~Sco indicate that it falls both in the extreme upper part of the instability strip for low-order $p$ mode oscillations (although far away from the known $\beta$~Cep variables clustered in the lower part of that instability strip; e.g. figs.~3,~4,~5 in \citeauthor{1999AcA....49..119P}~\citeyear{1999AcA....49..119P}; fig.~1 in \citeauthor{2011MNRAS.412.1814S}~\citeyear{2011MNRAS.412.1814S}) and in a confined region of the upper part of the HR diagram where the instability strip corresponding to $l=1-2$ $g$ mode oscillations reappears (the same modes encountered in slowly pulsating B-type stars, although V973~Sco is obviously far from the cluster of observationally confirmed SPB stars). However, regarding heat-driven pulsations, the matter of whether a given period $P$ might be due to pressure or gravity modes can even be more quantitatively evaluated with the associated value of the pulsation constant $Q=P\sqrt{(M/M_{\sun})/(R/R_{\sun})^3}$. The pulsation constant is of the order of $0.033-0.042$~d for the radial fundamental mode \citep{1973ApJ...179..235D,1974A&A....34..203L,1981ApJ...249..218F,2005ApJS..158..193S}, such that any value above this is indicative of gravity modes, while values below this correspond to pressure modes. For a $(45$M$_{\sun}, 36$R$_{\sun})$ star, $Q\simeq 0.03P$, meaning that heat-driven $p$ modes are expected to have periods shorter than $\sim$$1$~d for such stars, whereas periods longer than $\sim$$1$~d are likely to be those of $g$ modes. Nevertheless, heat-driven oscillations excited through the $\kappa$ mechanism operating in the iron opacity bump at $T\sim$$170$~kK are typically much more stable than the frequencies in set $\mathcal{S}_4$. Indeed, the spectrogram in Fig.~\ref{fig:V973Sco_BRITE_TFDiag} and better quantified through the analysis summarized in Table~\ref{tab:V973Sco_FreqAnalysis} says that, for instance, $\nu_{10}$ is only excited during epoch $\mathcal{E}_5$ for $\sim$$10$~d and is quasi-absent at other epochs. All these considerations lead us to the conclusion that caution must be exercised as for a strict interpretation of both the origin of the fact that $\nu_{11}\simeq2\nu_3$ and the origin of $\Delta\nu$, as, again, these could just be incidental owing to the high density of detected frequency peaks in the restricted $\sim$$[0.2;0.9]$~d$^{-1}$ region. 

The key information that needs to be considered here is the time-dependency of the different modes. The main sinusoidal signals composing the light variations vary in amplitude, and their appearance seems to occur in no particular organized manner, suggesting a stochastic triggering. After they are excited, they typically last about $\sim$$5-10$~days (this does not strictly apply to the signals affected by frequency resolution issues, as we already noted in detail in Section~\ref{sec:V973Sco_Results_Fourier}). Additionally, the presence of the bump in the regime $[0;0.9]$~d$^{-1}$ where the prominent signals are excited is indicative of stochastically excited modes with finite lifetimes, as is the case for e.g. stochastically excited solar-like oscillations \citep[e.g.][]{2010A&A...509A..77K}. All these behaviours (several excited frequencies in the range $[0;0.9]$~d$^{-1}$, clear bump in that frequency regime, time-dependency and finite lifetime of the modes) point towards three possible explanations for the dominant intrinsic light variations revealed by \emph{BRITE} in V973~Sco:

\begin{enumerate}[labelindent=4.0pt,leftmargin=*,label={(\roman*)}]
\item internal gravity waves (IGWs) excited from deep down at the level of the convective core--radiative envelope boundary \citep{2013ApJ...772...21R,2015ApJ...806L..33A,2017A&A...602A..32A};
\item gravity waves excited at the interface between a sub-surface convection zone and the radiative layer directly above it \citep{2009A&A...499..279C};
\item direct manifestation of turbulent convection in the upper part of a sub-surface convective zone.
\end{enumerate}

Regarding option (i), there have already been interpretations of the dominant stochastic light variations of three main-sequence and subgiant O-type stars observed by the \emph{COnvection ROtation and planetary Transits (CoRoT)} satellite (HD~46223 [O4\,V((f))], HD~46150 [O5\,V((f))z], and HD~46966 [O8.5\,IV]; \citeauthor{2011A&A...533A...4B}~\citeyear{2011A&A...533A...4B}; \citeauthor{2015ApJ...806L..33A}~\citeyear{2015ApJ...806L..33A}) and those of the O9.5\,Iab star HD~188209 observed by \emph{Kepler} \citep{2017A&A...602A..32A} as possibly arising from IGWs stochastically-triggered from the convective core. In order to further qualitatively compare the observed amplitude spectrum of V973~Sco with that of IGWs that could be excited from the convective core, we follow the same approach as that adopted by \citet{2015ApJ...806L..33A} and \citet{2017A&A...602A..32A}, which is based upon the two-dimensional hydrodynamical simulations performed by \citet{2013ApJ...772...21R}. We thus consider the case of the $3$M$_{\sun}$ non-rigidly rotating ZAMS model D11 of \citet{2013ApJ...772...21R} and apply a scaling factor to convert from tangential velocity to brightness amplitude, followed by a scaling of the frequencies to account for a TAMS supergiant. The validity and caveats in this simple scaling approach were addressed in detail by \citet{2015ApJ...806L..33A}. Evidently, the first requirement to be met in order to adopt this approach here is the similarity in the overall internal structure of a $3$M$_{\sun}$ ZAMS massive star and a star as massive and evolved as V973~Sco (i.e. convective core, radiative envelope). In spite of that being the case, it has been pointed out that the propagation of IGWs can be different depending on the stellar mass and evolutionary status. However, that effect might be lessened by the fact that the observable velocity and brightness amplitude spectra of IGWs are more influenced by the convective flux rather than the profile of the Br\"{u}nt--V\"{a}is\"{a}l\"{a} frequency \citep[equations 14 and 15]{2013ApJ...772...21R}. But in that regard, the total mass fraction of the convective core and the luminosity to mass ratio could be important parameters for the driving of the waves, and therefore not allow the scaling approach to yield optimal results. Additionally, as already noted by \citet{2015ApJ...806L..33A} and \citet{2017A&A...602A..32A}, obviously this approach only allows for a qualitative comparison of the overall behaviour of the simulated and observed periodograms, and does not allow a strict comparison of the excited frequencies one by one. Nevertheless, this approach currently remains the only viable one for V973~Sco until new hydrodynamical models of IGWs are available for the appropriate stellar mass range and age.

\begin{figure}
\includegraphics[width=8.4cm]{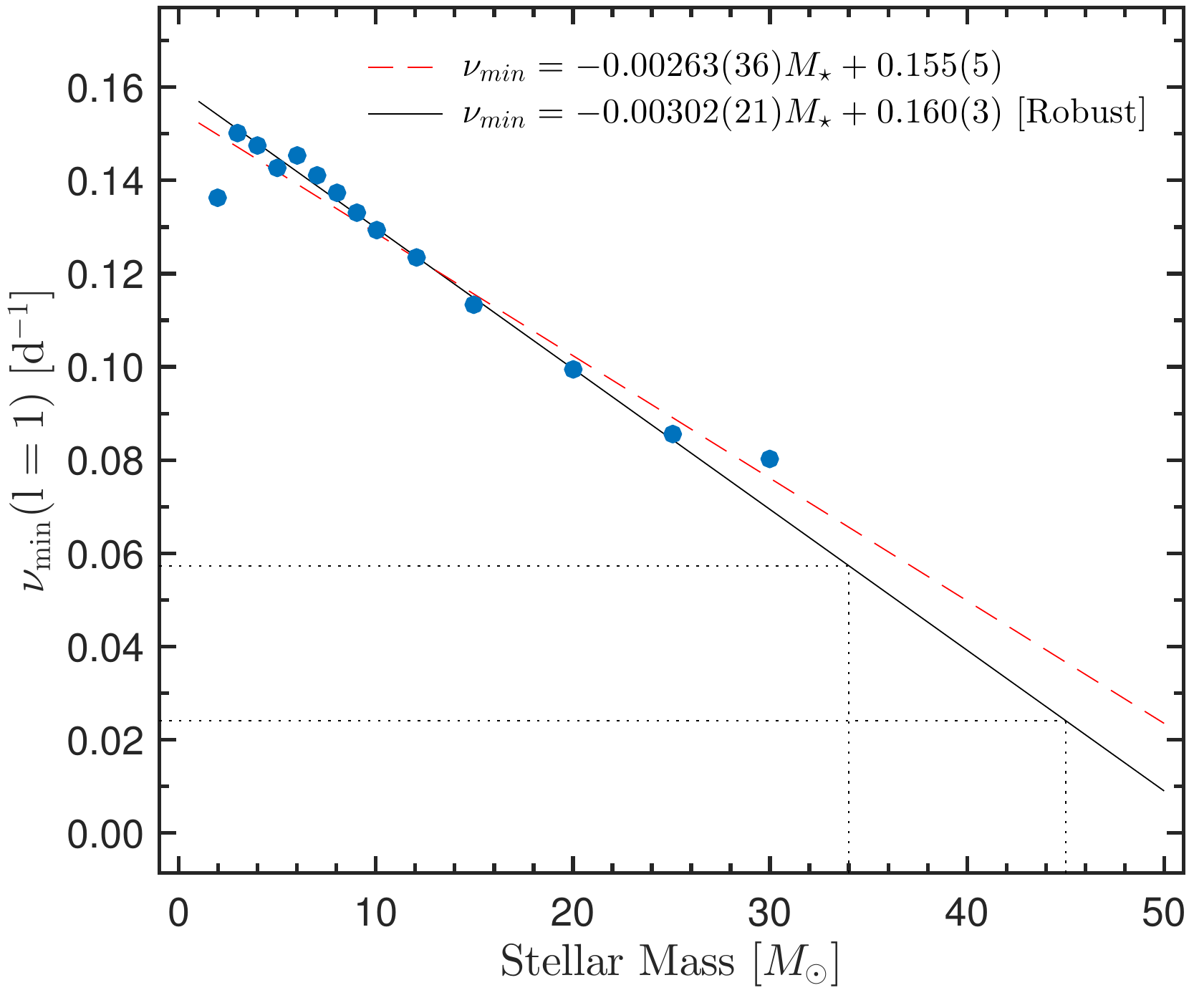}
 \caption{Trend followed by the minimum frequency for $l=1$ standing waves reported by \citet{2013MNRAS.430.1736S} as a function of stellar mass (pale blue dots), along with its linear least-square fit (dashed thin red line) and its robust linear least-square fit (solid black line). Dotted lines indicate the minimum frequency values corresponding to $M=34$M$_{\sun}$ and $M=45$M$_{\sun}$ as yielded by the robust fit.}
  \label{fig:V973Sco_Schiode13_Extrap}
\end{figure}

For the conversion from the actual simulated tangential velocity spectra to brightness amplitude spectra, we find a conversion factor of $19$~mmag/(km~s$^{-1}$) to be most appropriate to explain the observed amplitudes of the variability in V973~Sco. This value is about an order of magnitude larger than those found by \citet{2015ApJ...806L..33A} for the three main sequence and subgiant stars observed by \emph{CoRoT} (conversion factors in the range $1.5-3$~mmag/(km~s$^{-1}$)). We note that if the same procedure was to be applied to the cases of the amplitude spectra of the \emph{Kepler} observations of HD 188209 \citep[O9.5\,Iab;][]{2017A&A...602A..32A} and the residual light variations of $\zeta$~Pup as observed by \emph{BRITE} \citep[O4\,I(n)fp; and `residual' means free of any signs of the $1.78$-d rotation of the star;][]{2018MNRAS.473.5532R}, the conversion factors would be of the order of $4$~mmag/(km~s$^{-1}$) for both stars. In that regard, V973~Sco appears to be showing higher IGW brightness variability. Obviously, more observations are needed to confirm this hypothesis and to check whether there is any correlation with spectral type.

In terms of the scaling in frequencies, \citet{2015ApJ...806L..33A} and \citet{2017A&A...602A..32A} used the results of \citet[][their table~1]{2013MNRAS.430.1736S} to find the appropriate scaling factor between the frequencies of $l=1$ gravity waves of a $3$M$_{\sun}$ ZAMS star and those for stars having the mass and evolutionary status of their targets. In the case of V973~Sco, \citet{2005A&A...436.1049M} quote a typical mass of $\sim$$34$M$_{\sun}$ for an O8.5 supergiant\footnote{\citet{2005A&A...436.1049M} only consider O stars divided up into luminosity classes V, III and I. They do not distinguish luminosity sub-classes Ib, Iab, Ia and Ia+. V973 Sco being of luminosity class Ia, it is brighter than the average luminosity class I considered by \citet{2005A&A...436.1049M}, thus expected to be more massive than the $\sim$$34$M$_{\sun}$ mass they quote for luminosity class I O-type stars.}, whereas \citet{2009A&A...503..985C} estimated a current mass in the range $\sim$$35-45$M$_{\sun}$ for V973~Sco. Since the results reported by \citet{2013MNRAS.430.1736S} do not go beyond $30$M$_{\sun}$, we extrapolate their values by performing a first order (linear) fit to the trend followed by the minimum frequencies for dipole gravity waves as a function of stellar masses, as shown in Fig.~\ref{fig:V973Sco_Schiode13_Extrap}. Adopting the robust fit to the trend, we find a scaling factor of $0.16$ for a $45$M$_{\sun}$ star and $0.38$ in the case of a $34$M$_{\sun}$ star.

\begin{figure*}
\includegraphics[width=18cm]{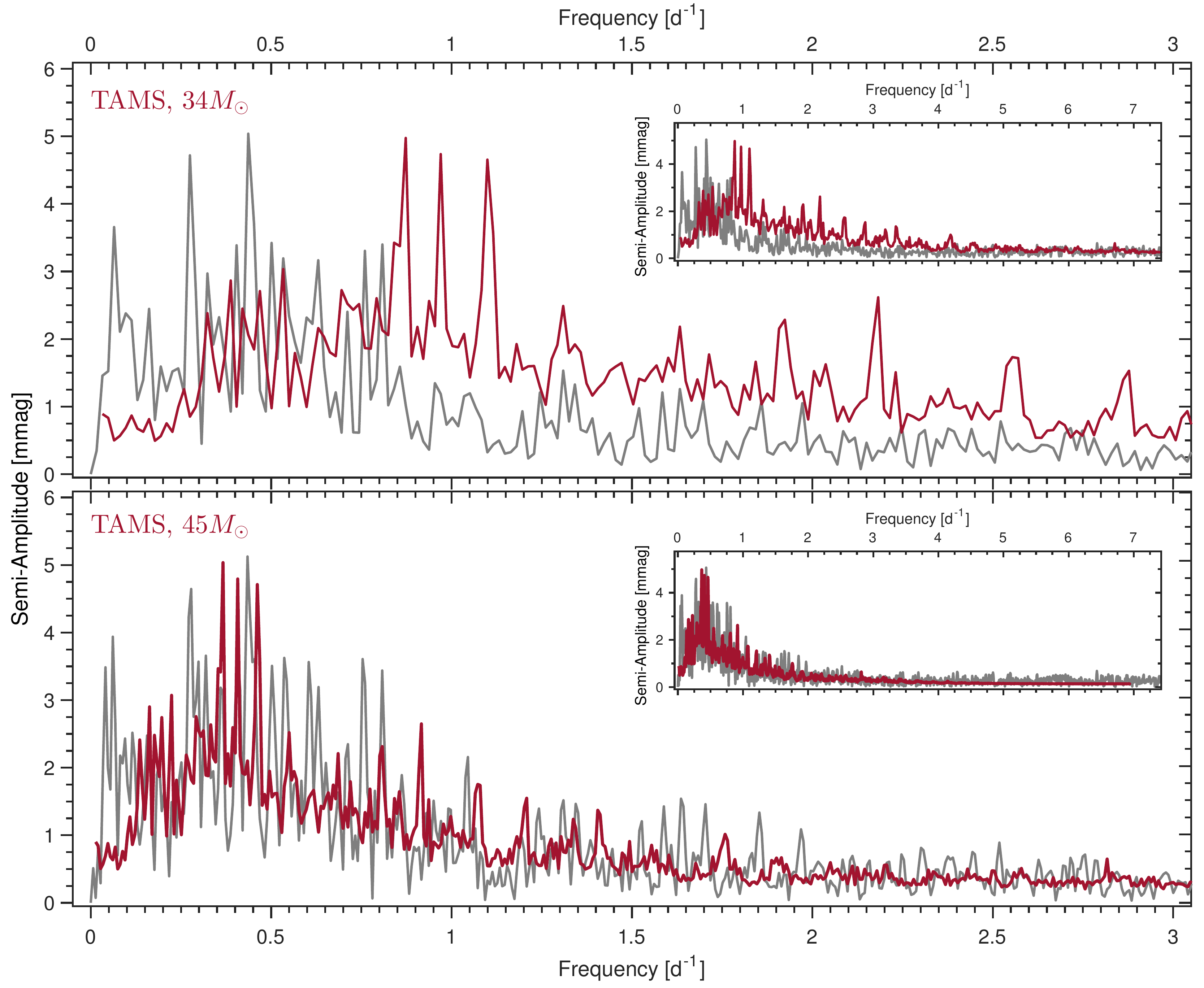}
 \caption{The observed amplitude spectrum of the \emph{BHr} light curve of V973~Sco (gray) compared to the emergent amplitude spectrum from 2D HD simulations of IGWs triggered from deep in the convective core (red), corresponding to model D11 of \citet[][]{2013ApJ...772...21R} but properly scaled in frequency to account for the status of V973~Sco as a supergiant and two possibilities for its current mass ($\sim$$34$M$_{\sun}$ according to \citeauthor{2005A&A...436.1049M}~\citeyear{2005A&A...436.1049M}; $\sim$$35-45$M$_{\sun}$ according to \citeauthor{2009A&A...503..985C}~\citeyear{2009A&A...503..985C}). Clearly the $45$M$_{\sun}$ model corresponds better to the behaviour of the observed amplitude spectrum. Note that in order to perform a fair comparison between the observed and the modelled amplitude spectra, the former had to be re-evaluated with the same frequency resolution as the latter.}
  \label{fig:V973Sco_BRITE_DFTvsIGWsHD}
\end{figure*}

Figure~\ref{fig:V973Sco_BRITE_DFTvsIGWsHD} depicts the outcome of the comparison process, clearly indicating that the $45$M$_{\sun}$ model is favoured as its broad bump in the amplitude spectrum and the associated range of most excited frequencies match well with the ones observed in V973~Sco. We emphasize here that these simulations do not allow a strict comparison of the frequency peaks one by one, but serve as a qualitative comparison of the global shape, trend, and range of excited frequencies of the periodograms.

Concerning options (ii) and (iii), we note that further investigations need to be conducted on the possibility of IGWs that could be excited from a potential subsurface convection zone \citep[][]{2009A&A...499..279C}, and/or whether the effects of turbulent convection in such a subsurface convection zone could be visible up to the level of the photosphere. Adding to that is the fact that, at the moment, it is unclear whether the velocity/brightness amplitude spectrum due to breaking, stochastically-generated IGW \citep[as simulated by][]{2013ApJ...772...21R} is fundamentally different from the spectrum of turbulent convection. Nevertheless, in both cases evidently the observable effects could become more important the closer to the surface the convective layer is located. In that regard, the current status of V973~Sco turns out to be particularly interesting. According to the findings of \citet[][]{2009A&A...503..985C}, the current mass of V973~Sco is of the order of $35-45$M$_{\sun}$, which corresponds to an initial mass of about $60$M$_{\sun}$ and an age of $\sim$$2.7$~Myr. This means that V973~Sco would be the ideal archetype of the M$_{\rm ZAMS}=60$M$_{\sun}$ model considered by \citet[][their figs. 1 and 3]{2009A&A...499..279C}. Particularly, the age estimate of $\sim$$2.7$~Myr corresponds to a stage where a convection layer due to the iron opacity bump at $T\simeq$$170$~kK plunges further down into the star (therefore, perhaps better qualified as `intermediate' convective zone rather than `subsurface' convective zone), while a subsurface convective layer due to the helium opacity peak at $T\simeq$$50$~kK appears at $\sim0.12$R$_{\sun}$ below the photosphere \citep[fig. 3 of][]{2009A&A...499..279C}. These considerations emphasize the need for in-depth theoretical investigations on options (ii) and (iii).

\section{Conclusions and future work}
\label{sec:V973Sco_Discussion}

The \emph{BRITE} time-dependent photometric observations of V973~Sco that we have analyzed in this investigation are the longest and most intensive observations of V973~Sco to date, and revealed for the first time the unambiguous presence of a stochastically-generated intrinsic variability at the photosphere of the star. Stochastic triggering of IGWs from the convective core yields a brightness amplitude spectrum that qualitatively matches the observed amplitude spectrum of the light variations of V973~Sco, especially the bump where the dominant modes are located. The evident next path to be explored from the theoretical standpoint is the investigation of the possibility of stochastic excitation of gravity waves from a subsurface convection layer, as well as an in-depth investigation of the observable characteristics of turbulent convective motions in such subsurface convective regions.

On the observational side, despite the fact that for V973~Sco these observations are unprecedented in terms of their quality, cadence, and time span, we emphasize the (obvious) need for observations on a much longer time base to better quantify the characteristics of the modes involved in these IGWs, especially their lifetimes, the frequency resolution of the current time series photometric observations being still insufficient to clearly distinguish some signals, as we pointed out in Section~\ref{sec:V973Sco_Results_Fourier}. Additionally, future observations of V973~Sco need to explore the connection between these IGWs appearing at the photosphere and wind clumping activity. The four-year long photometric and spectroscopic multi-site campaign conducted by \citet{2017A&A...602A..32A} on the late O-type supergiant HD 188209 revealed that the stellar photospheric variability due to IGWs propagates into the stellar wind. The six-month long \emph{BRITE} observations of the early O-type supergiant $\zeta$~Puppis, complemented by contemporaneous ground-based multi-site spectroscopic monitoring of the He~{\sc ii}~$\lambda4686$ wind-sensitive emission line of the star, revealed that large-scale wind structures are driven by bright photospheric spots and that the inner wind clumping activity is triggered by an unknown stochastic mechanism operating at the photosphere and manifesting itself as light variations of stochastic nature \citep{2018MNRAS.473.5532R}. The most probable scenario that can be drawn from those two recent investigations is that stochastically excited IGWs seen at the stellar photosphere induce clumping at the very base of the wind. But, clearly, more long-term, multi-wavelength observations (e.g. long-duration, continuous, contemporaneous optical and X-ray monitoring -- the X-ray domain being useful to probe shocks that could be related to wind clumps) on a relatively large number of O-type stars is needed in order to obtain a definitive, complete picture on this point. This has already been initiated on $\zeta$~Pup, for which a $840$~ks \emph{Chandra} X-ray observing run will be executed (PI: Wayne Waldron) in parallel with ground-based spectroscopy and monitoring by \emph{BRITE-Constellation}.

\section*{Acknowledgements}

AFJM acknowledges support from the Canadian Space Agency (CSA), the Natural Sciences and Engineering Research Council of Canada (NSERC), and the Fonds de Recherche du Qu\'{e}bec - Nature et Technologies (FRQNT). APo acknowledges NCN grant  2016/21/D/ST9/00656 and used the infrastructure supported by POIG.02.03.01-24-099/13 grant: GCONiI--Upper-Silesian Center for Scientific Computation. APi acknowledges support from the NCN grant no. 2016/21/B/ST9/01126. The Polish contribution to the \emph{BRITE} mission is supported by a SPUB grant of the Polish Ministry of Science and Higher Education (MNiSW). GH acknowledges support from the Polish NCN grant 2015/18/A/ST9/00578. GAW acknowledges Discovery Grant support from NSERC. TR
was supported in part by science funding related to the \emph{BRITE-Toronto} mission, awarded by the CSA to GAW on behalf of the Canadian \emph{BRITE} team. KZ acknowledges support by the Austrian Space Application Programme (ASAP) of the Austrian Research Promotion Agency (FFG).








\appendix

\section{\emph{BRITE} photometry of V973~Sco: decorrelation with respect to trends of instrumental origin}
\label{sec:V973Sco_Decorrelation}

\begin{figure}
\includegraphics[width=8.4cm]{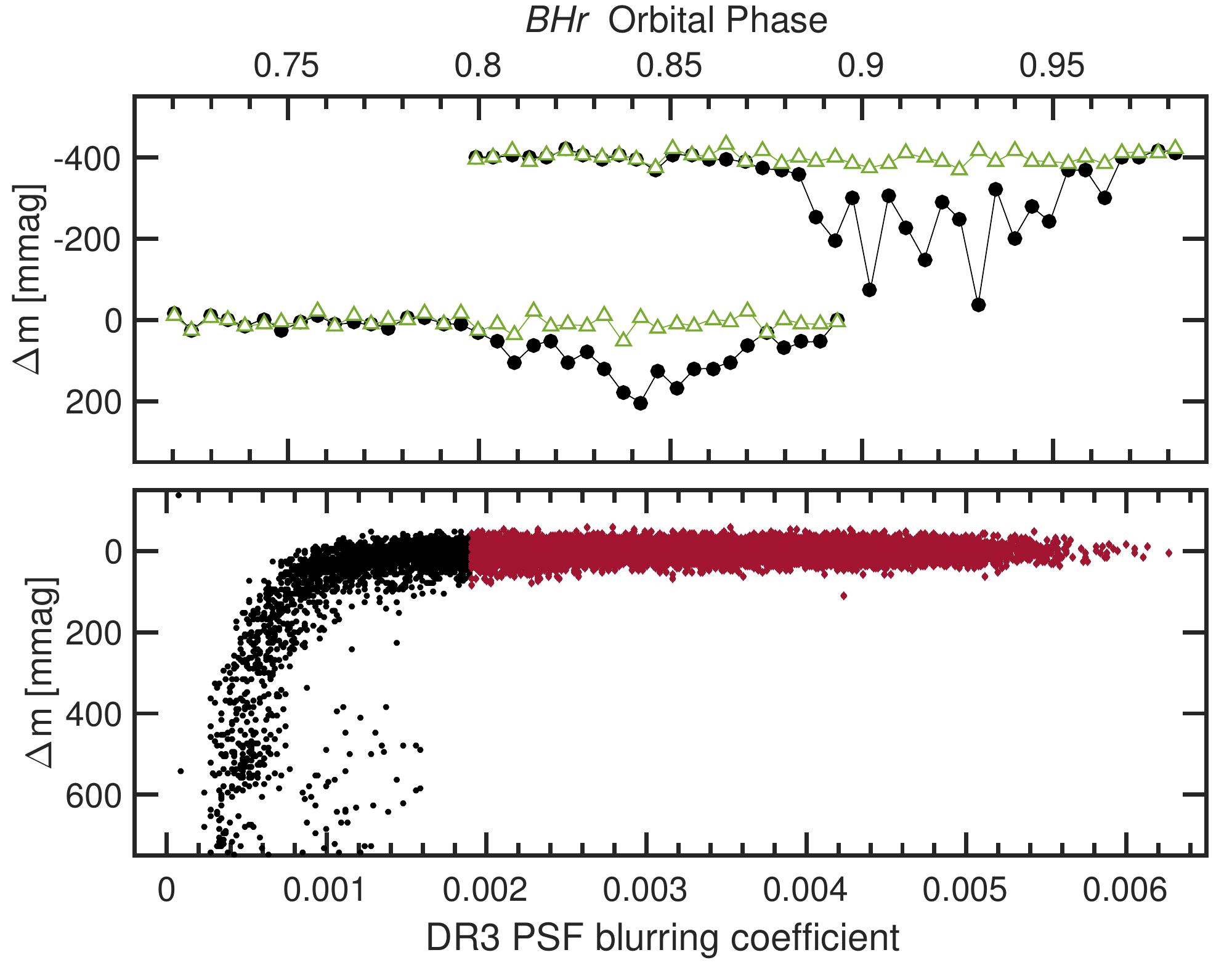}
 \caption{Illustration of the occurence of light curve anomalies corresponding to a strong blurring of the PSF in the \emph{BHr} observations of V973~Sco released in the ``DR3'' format. \emph{Top:} Sample time series taken within two different satellite orbits, respectively centered around HJD~$2457204.84851$ and HJD~$2457218.27183$. The second set of observations is offset by $0.4$~mag for better visibility. Filled circles correspond to the officially released data, showing the anomalies, while open triangles correspond to the data that we re-extracted using the most updated version of the reduction pipeline. \emph{Bottom:} A view of the officially released data as a function of the ``DR3'' PSF blurring coefficient. Black points represent the entirety of the data, showing strong flux loss for the most blurred observations, whereas red points correspond to the measurements that would remain when cutting off observations with blurring coefficient values lower than $1.9\times 10^{-3}$.}
  \label{fig:V973Sco_BRITE_blurring_DR3}
\end{figure}

An initial version of the \emph{BRITE} observations of V973~Sco was extracted from raw images to a time series of flux measurements (corrected for intrapixel variations) using the standard pipeline for \emph{BRITE} data reduction described in \citet{2017A&A...605A..26P}, and was officially released under format ``DR3''\footnote{see Table A.1 in \citeauthor{2017A&A...605A..26P}~\citeyear{2017A&A...605A..26P} for the different existing versions of the release formats of \emph{BRITE} light curves.}. That format allows for the correction for possible trends of instrumental origin related to the location $(x_{\rm cen};y_{\rm cen})$ of the centroid of the stellar point spread function (PSF) on the chip, variations in the temperature $T_{\textsc{ccd}}$ of the detector, the satellite orbital phase $\phi_{\rm orb}$ (from the time column and with $P_{\rm orb} = 97.0972$~min), blurring of the stellar PSF that might be due to imperfect satellite stability (characterized by a coefficient named ``PSFC''), and effects of random telegraph signal (RTS - indicated by a coefficient named ``RTSC''). The most prominent instrumental effect that we noticed in that initial version of the raw \emph{BRITE} light curve of V973~Sco is a sudden drop in flux in some of the orbits, corresponding to a strong blurring of the stellar PSF as indicated by the behaviour of the ``PSFC'' coefficient (Fig.~\ref{fig:V973Sco_BRITE_blurring_DR3}). This effect could arise from some scattered light coming to the star-tracker, or other issues related to the orientation of the satellite relative to the Sun, the Earth and the Moon. Furthermore, as depicted in Fig.~\ref{fig:V973Sco_BRITE_blurring_DR3}, this effect does not necessarily happen at the same orbital phase, such that decorrelation with respect to $\phi_{\rm orb}$ does not account for this anomaly. Unfortunately, the most appropriate way of dealing with this is to clip off measurements showing strong blurring effect, namely measurements with low ``PSFC'' values. In that regard, the choice of an appropriate threshold remains an issue and can even be qualified as arbitrary. Adopting a trial-and-error approach (by testing a range of ``PSFC'' value thresholds and inspecting the resulting remaining measurements taken within each orbit in the time series), we determined a value of $1.9\times10^{-3}$ to be a good limit to adopt (Fig.~\ref{fig:V973Sco_BRITE_blurring_DR3}). However, this obviously results in a sharp cut and removes some points that are only slightly blurred but not affected by a strong flux loss. Given the fact that more recent formats of \emph{BRITE} data releases (``DR4'' and ``DR5'') contain an additional coefficient to better characterize blurring effects, we decided to re-extract the \emph{BHr} observations of V973 Sco using the most updated, improved version of the reduction pipeline for observations performed in chopping mode. This allowed us to extract:
\begin{enumerate}[labelindent=4.0pt,leftmargin=*]
\renewcommand\labelenumi{(\roman{enumi})}
\item four coefficients to characterize the blurring effect (hereafer PSFC$_{j=1...4}$);
\item the number of pixels $N_{\rm pix}$ considered in the stellar PSF;
\item an estimate of intensity offsets $I_{\rm off}$ within the custom aperture enclosing the target;
\item an estimate of the background level $B_{\rm diff}$ in difference images.
\end{enumerate}

\begin{figure}
\includegraphics[width=8.4cm]{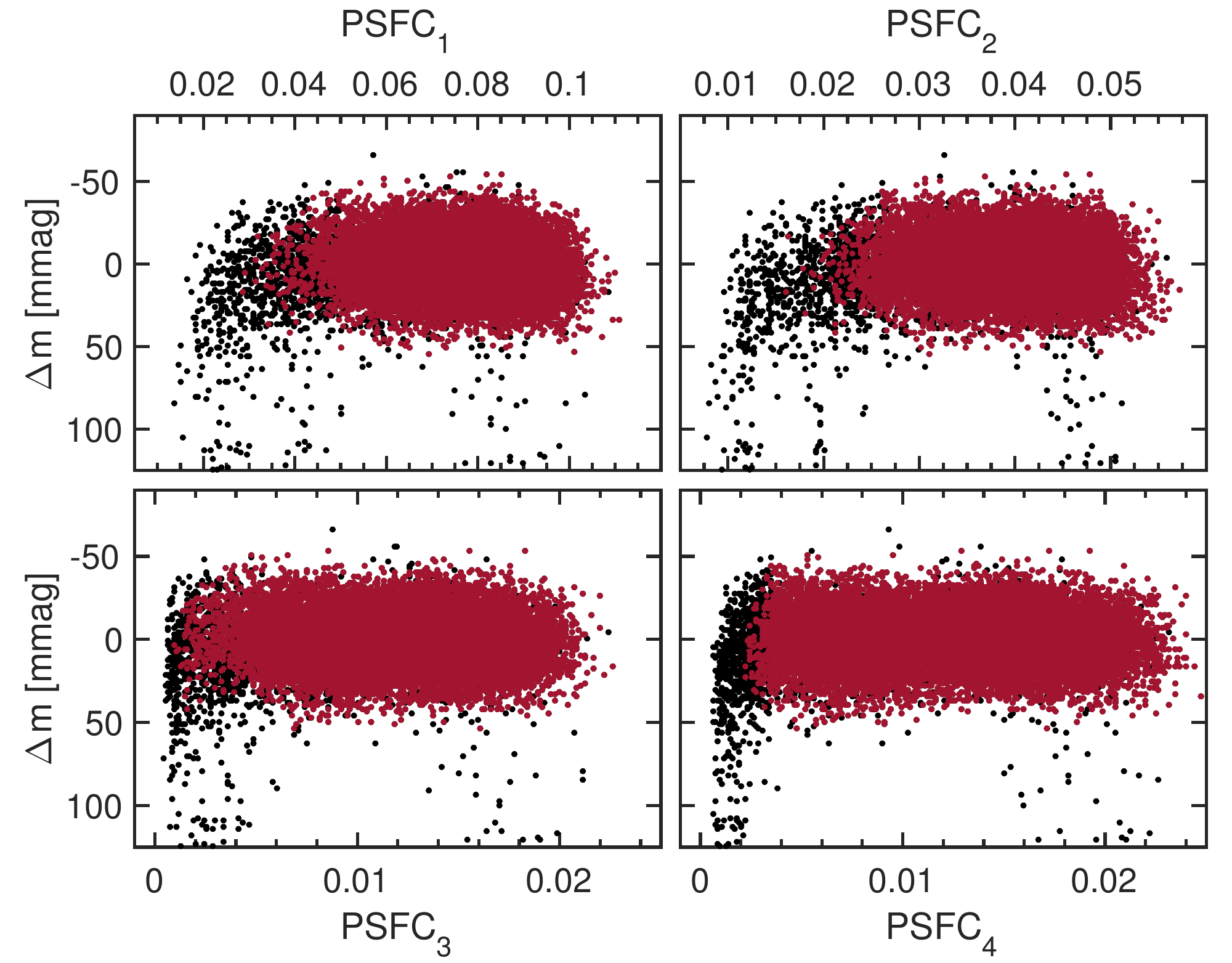}
 \caption{Newly extracted raw \emph{BHr} light cure of V973~Sco viewed as a function of the four blurring coefficients PSFC$_{j=1...4}$, showing fairer results out of the thresholding process. Black points are the raw data, whereas red points represent the remaining points after clipping off measurements that do not satisfy the condition $\mathbf{(C)}$ defined in the text.}
  \label{fig:V973Sco_BRITE_blurring_new}
\end{figure}

The motivations for considering blurring effects in \emph{BRITE} observations were presented in detail by \citet{2017A&A...605A..26P}. The parameter PSFC$_2$ is the same energy-based indicator ``PSFC1'' defined in equation B.1 by \citet{2017A&A...605A..26P} for ``DR4'' and ``DR5'' formatted data. The parameters PSFC$_3$ and PSFC$_4$ are the decomposed versions of the correlation-based indicator ``PSFC2'' introduced in ``DR4'' and ``DR5'' formatted data \citep[equation B.2 of][]{2017A&A...605A..26P}, now separately measuring blurring effects in the vertical direction (PSFC$_3$) and in the horizontal direction (PSFC$_4$). The only genuinely new blurring coefficient that we have also assessed here is PSFC$_1$, which simply tracks the ratio of the intensity of the brightest pixel in the stellar PSF to the integrated intensity over the stellar PSF.

Another useful parameter is the estimate of possible intensity offsets within the apertures ($I_{\rm off}$). For a given observation in which a specific aperture has been assigned to the target, $I_{\rm off}$ considers only the pixels in that aperture and evaluates the sum of their intensity differences between the next observation and the previous one. This is a good proxy for measuring a possible increase of background light within the aperture. We note that this method for estimating $I_{\rm off}$ is only possible in the chopping mode of observation as a result of the fact that the position of the stellar PSF in the current image is shifted in the $x$ direction with respect to the PSF positions in the precedent and subsequent images. 

Finally, the parameter $B_{\rm diff}$ is exactly the same as the parameter ``APER0'' described by \citet{2017A&A...605A..26P} for data in ``DR4'' or ``DR5'' formats, and could be useful in that it accounts for possible occasional intensity gradients in background between exposures, an effect that could induce some larger scatter within some orbits.

As readily visible in the top panel of Fig.~\ref{fig:V973Sco_BRITE_blurring_DR3}, the new reduction of the \emph{BRITE} light curve of V973~Sco is better than the initial reduction, as some of the anomalies related to the blurring of the PSF are mitigated. The reason for this improvement is chiefly an improved best aperture selection in the latest version of the pipeline, essentially resulting from the fact that corrections for intrapixel variations are now done separately for the two sets of data in the two chopping positions (instead of globally as done in the case of ``DR3'' formatted data).

In the newly reduced time series of V973~Sco, the first step that we performed was the removal of obvious outliers. In that step, we ended up with the following condition for a measurement to be considered for further processing:
\begin{equation*}
\mathbf{(C)} \begin{cases}
~(\text{PSFC}_1>0.05) \vee (\text{PSFC}_2>0.03) \vee (\text{PSFC}_3>0.006) \\
~~~~~~~~~~~~~~~~~~~~~\vee (\text{PSFC}_4>0.0035)\\
~|B_{\rm diff}-23| < 27 \\
~|I_{\rm off}| < 1200\\
~\left(|x_{\rm cen}+8.05| < 2.55\right) \vee (|x_{\rm cen}-11.00| < 2.00 )\\
~|y_{\rm cen}| < 3.50,
\end{cases}
\end{equation*}   
with `$\vee$' denoting the inclusive disjunction. Fig.~\ref{fig:V973Sco_BRITE_blurring_new} illustrates the resulting remaining measurements, particularly showing the capability of the four coefficients in keeping some of the measurements having low PSFC$_{j=1...4}$ values but not affected by strong flux deficit. At this stage of the light curve cleaning procedure we measure $\sigma_{\rm rms}\simeq2.28$~mmag for the rms value of the mean standard deviations per orbit.

The next step consists in removing possible trends of instrumental origin affecting the flux measurements by exploring their behaviour as a function of the eleven parameters available, i.e. $(x_{\rm cen};y_{\rm cen})$, $T_{\textsc{ccd}}$, PSFC$_{j=1...4}$, $N_{\rm pix}$, $I_{\rm off}$, $B_{\rm diff}$, and $\phi_{\rm orb}$. The multivariate nature of the problem and our aim to concurrently describe the possible correlations between the different variables naturally calls for an approach involving principal component analysis (PCA). In fact, our time series of flux measurements of V973 Sco taken along with the estimates of the aforementioned eleven parameters constitute a series of $n=18841$ measurements of $m=12$ possibly dependent variables that can be regarded as an $n\times m$ matrix $\mathcal{M}$, each line representing the coordinates of a data point in an $m$-dimensional space formed by each of the original variables. Then we can use PCA to find the new coordinates of  the set of measurements in a new orthogonal basis, the axes of which are directed by the eigenvectors of the covariance matrix $\mathcal{C}$ of $\mathcal{M}$. However, the \emph{BHr} observations of V973~Sco were performed in chopping mode in the $x$ direction, obviously resulting in a globally bimodal distribution of the values of $x_{\rm cen}$, such that it would be meaningless to treat this parameter globally together with the other parameters using PCA. Moreover, PCA only accounts for linear correlations, which could pose an issue if some of the trends of instrumental origin are not necessarily linear as in the case of other stars observed by \emph{BRITE} \citep[see e.g. fig. A.5 in][]{2016A&A...588A..55P}. In consideration of these two caveats, we adopted the following three-step strategy:

\begin{enumerate}[labelindent=4.0pt,leftmargin=*,label={(\arabic*)}]
\item separate the data in two sets corresponding to the two distributions of PSF centroid $x$ position;
\item perform PCA on these two sets;
\item recombine the resulting corrected sets and adopt the method described by \citet{2016A&A...588A..55P} to correct for non-linear trends in the recombined data.
\end{enumerate}

\begin{figure*}
\includegraphics[width=18cm]{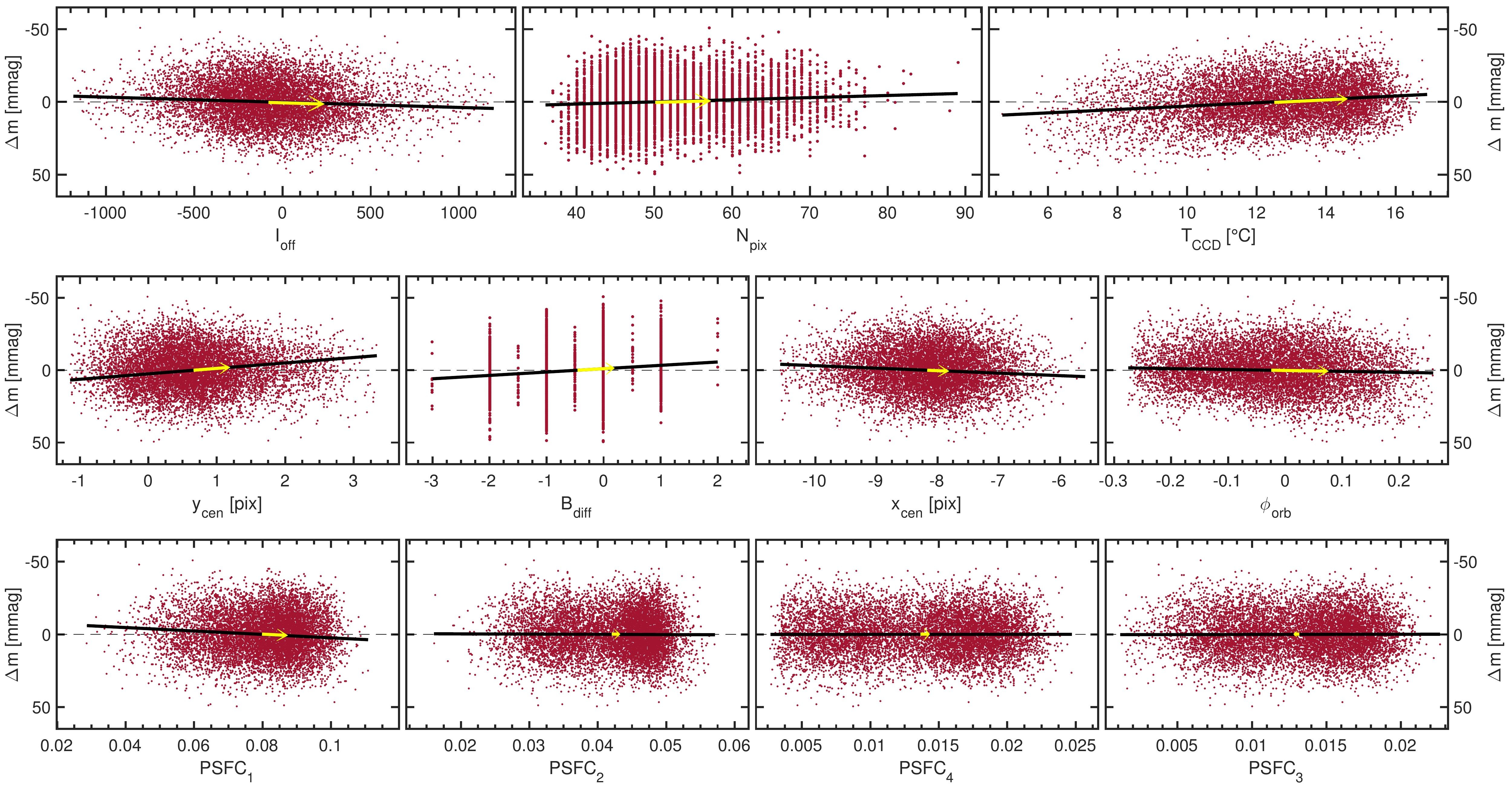}
 \caption{Outcome of PCA on the \emph{BRITE} observations of V973~Sco for one of the centroid $x$ position. Each panel illustrates the projection of the observations onto the plane of flux measurements versus one of the eleven instrumental parameters (red points), along with the corresponding direction of linear correlation/anti-correlation found by PCA (black line) as directed by the projection of the corresponding eigenvector (yellow arrow) of the covariance matrix onto that plane.}
  \label{fig:V973Sco_BRITE_decorr_PCA_X1R}
\end{figure*}

\begin{figure*}
\includegraphics[width=18cm]{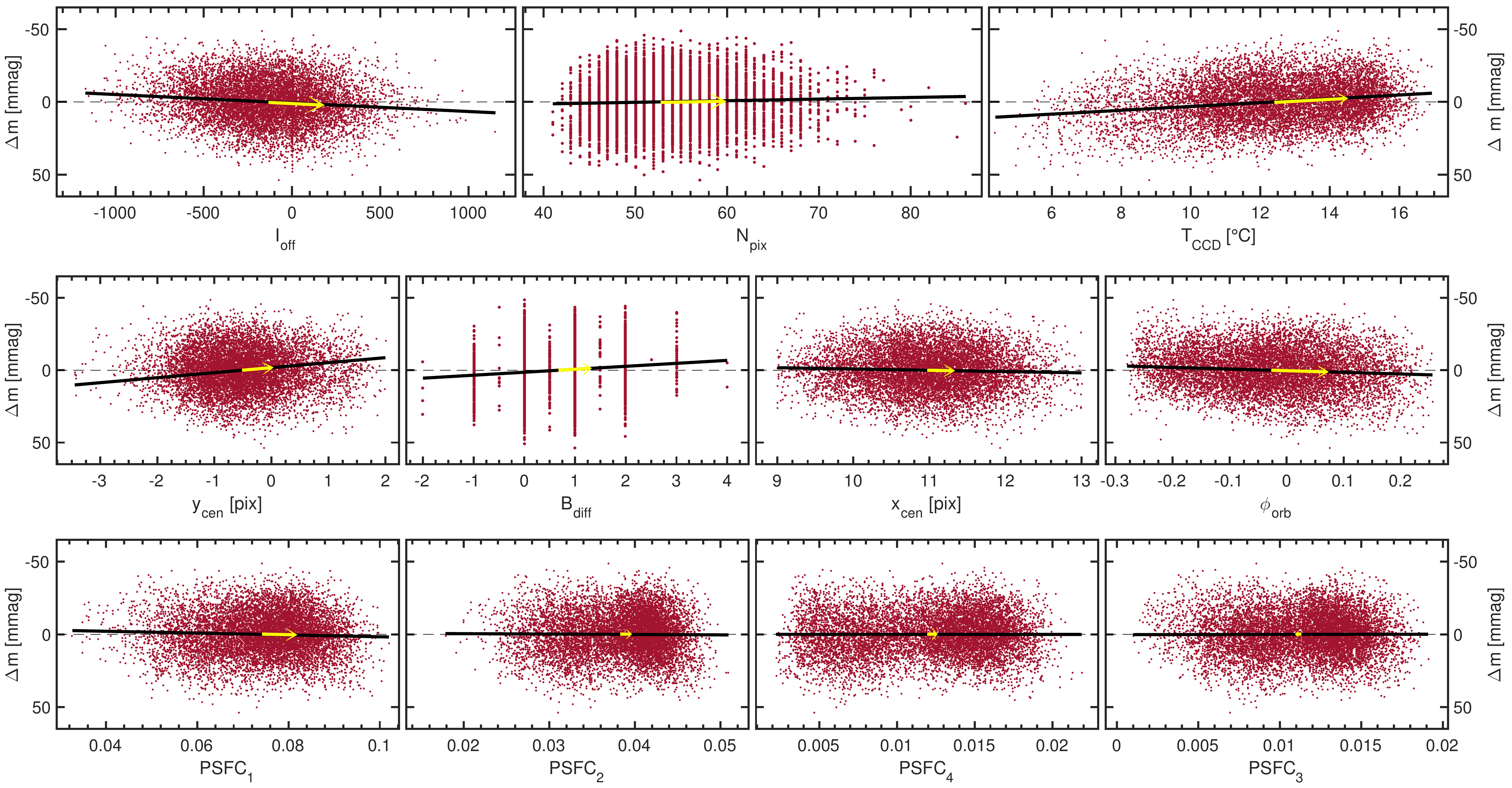}
 \caption{Same as Fig.~\ref{fig:V973Sco_BRITE_decorr_PCA_X1R} but for the part of the observations acquired at the other centroid $x$ position.}
  \label{fig:V973Sco_BRITE_decorr_PCA_X2R}
\end{figure*}

\begin{figure}
\includegraphics[width=8.4cm]{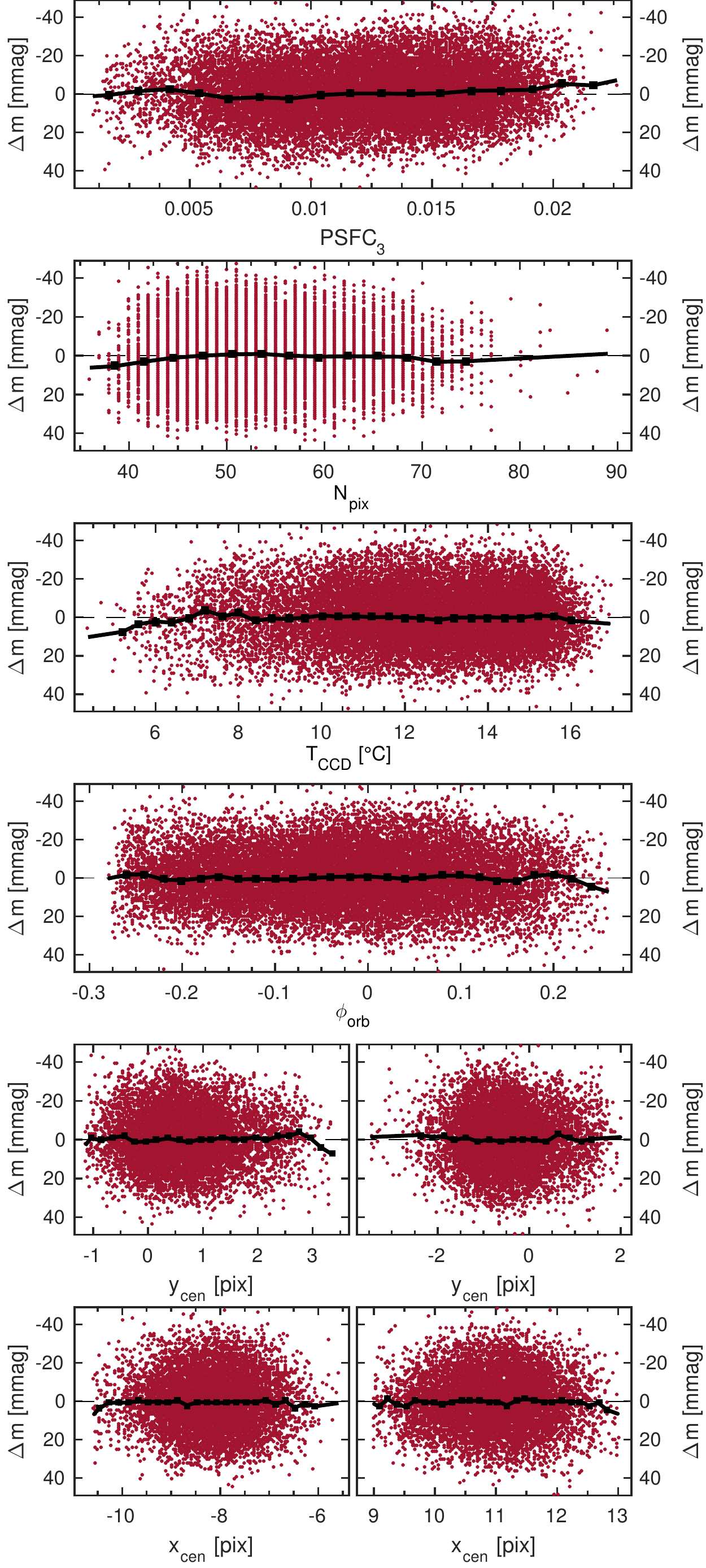}
 \caption{Correction of the remaining non-linear trends in the \emph{BRITE} observations of V973~Sco found as a function of some of the instrumental parameters.}
  \label{fig:V973Sco_BRITE_decorr_XYPhiTNP3}
\end{figure}

We note that, steps (2) and (3) could be substituted by a single step involving nonlinear PCA \citep[NLPCA; e.g. ][]{1991Kramer,1996DongMcAvoy,2012Scholz}. In that case, NLPCA would be applied on the two sets, which are eventually recombined after the nonlinear trends are corrected. However, after testing an implementation of NLPCA based on an auto-associative neural network \citep{2012Scholz} on the \emph{BRITE} observations of V973~Sco,  we found that, not only is it computationally expensive, but it also yielded  unsatisfactory results in describing some highly non-linear trends in the data (e.g. the behaviour of flux as a function of orbital phase). Thus, we decided to adopt the strategy involving the aforementioned three steps. Figs.~\ref{fig:V973Sco_BRITE_decorr_PCA_X1R}~and~\ref{fig:V973Sco_BRITE_decorr_PCA_X2R} depict the outcome of the PCA on the two separate data sets, while Fig.~\ref{fig:V973Sco_BRITE_decorr_XYPhiTNP3} plots the non-linear trends detected and corrected in the recombined data after the PCA. Finally, we perform a global $4\sigma$ clipping to the detrended data, followed by a $2\sigma$ orbital clipping, and omit any orbit containing less than five points. The properties of the final light curve at post-decorrelation stage are summarized in the last four entries of Table~\ref{tab:V973Sco_Obs_BRITE_Log} and the light curve binned over each \emph{BHr} orbit illustrated on Fig.~\ref{fig:V973Sco_BRITE_lcs_full}.

\bsp	
\label{lastpage}
\end{document}